\documentclass[usenatbib,useAMS]{mn2e}

\usepackage{wasysym}
\usepackage{graphicx}
 
\begin{document}

\label{firstpage}

\title[Transition from star clusters to galaxies]{From star clusters to dwarf galaxies: The properties of dynamically hot stellar systems}

\author[Dabringhausen et al.]{
J. Dabringhausen$^{1}$ \thanks{E-mail: joedab@astro.uni-bonn.de},
M. Hilker$^{2}$ \thanks{mhilker@eso.org} and
P. Kroupa$^{1}$ \thanks{pavel@astro.uni-bonn.de}\\
$^{1}$ Argelander-Institut f\"ur Astronomie, Universit\"at Bonn, Auf dem H\"ugel 71, 53121 Bonn, Germany\\
$^{2}$ European Southern Observatory, Karl-Schwarzschild-Stra\ss e 2, 85748 Garching bei M\"unchen, Germany}

\pagerange{\pageref{firstpage}--\pageref{lastpage}} \pubyear{2007}

\maketitle

\begin{abstract}
Objects with radii of $10 \, \rmn{pc}$ to $100 \, \rmn{pc}$ and masses in the range from $10^6 \, \rmn{M}_{\odot}$ to $10^8 \, \rmn{M}_{\odot}$ have been discovered during the past decade. These so-called ultra compact dwarf galaxies (UCDs) constitute a transition between classical star clusters and elliptical galaxies in terms of radii, relaxation times and $V$-band mass-to-light ratios. Using new data, the increase of typical radii with mass for compact objects more massive than $10^6 \, \rmn{M}_{\odot}$ can be confirmed. There is a continuous transition to the typical, mass-independent radii of globular clusters (GCs). It can be concluded from the different relations between mass and radius of GCs and UCDs that at least their evolution must have proceeded differently, while the continuous transition could indicate a common formation scenario. The strong increase of the characteristic radii also implies a strong increase of the median two-body relaxation time, $t_{\rmn{rel}}$, which becomes longer than a Hubble time, $\tau_{\rmn{H}}$, in the mass interval between $10^6 \, \rmn{M}_{\odot}$ and $10^7 \, \rmn{M}_{\odot}$. This is also the mass interval where the highest stellar densities are reached. The mass-to-light ratios of UCDs are clearly higher than the ones of GCs, and the departure from mass-to-light ratios typical for GCs happens again at a mass of $\approx 10^6 \, \rmn{M}_{\odot}$. Dwarf spheroidal galaxies turn out to be total outliers compared to all other dynamically hot stellar systems regarding their dynamical mass-to-light ratios. Stellar population models were consulted in order to compare the mass-to-light ratios of the UCDs with theoretical predictions for dynamically unevolved simple stellar populations (SSPs), which are probably a good approximation to the actual stellar populations in the UCDs. The SSP models also allow to account for the effects of metallicity on the mass-to-light ratio. It is found that the UCDs, if taken as a sample, have a tendency to higher mass-to-light ratios than it would be expected from the SSP models assuming that the initial stellar mass function in the UCDs is the same as in resolved stellar populations. This can be interpreted in several ways: As a failure of state-of-the-art stellar evolution and stellar population modelling, as a presence of dark matter in UCDs or as stellar populations which formed with initial stellar mass functions different to the canonical one for resolved populations. But it is noteworthy that evidence for dark matter emerges only in systems with $t_{\rmn{rel}} \apprge \tau_{\rmn{H}}$.
\end{abstract}


\section{Introduction}
\label{Introduction}
Star clusters can be defined as stellar population with a median two-body relaxation time, $t_{\rmn{rel}}$, shorter than a Hubble time, $\tau_{\rmn{H}}$, while galaxies would have $t_{\rmn{rel}} > \tau_{\rmn{H}}$ \citep{Kro1998}. The dynamical evolution of the former is well described by pure Newtonian dynamics, while for the successful representation of the latter either a significant amount of dark matter (DM) is required for Newtonian gravity to remain valid, or modified gravity needs to be invoked. By moving from two-body relaxation dominated systems to such where two-body relaxation plays no role, we thus observe the appearance of fundamentally new physics. A transition class of objects between classical star clusters and galaxies may shed insights to the possible nature of the deviant dynamics apparent on galaxy scales.

It has been almost 10 years since \citet{Hil1999} and \citet{Dri2000} discovered these transition objects in the Fornax galaxy cluster. With apparent $V$-band magnitudes of $\apprle 19.5 \, \rmn{mag}$ at that distance, they can in principle be detected without difficulty. However, they cannot be discriminated from point sources with ground-based telescopes, except with the ones with the highest currently available resolutions. Because of this combination of small extension and high brightness they were usually thought to be foreground stars. Only a radial velocity survey of \emph{all} objects with a certain brightness in an area around the central galaxy of the Fornax cluster was able to reveal their membership to that galaxy cluster. \citet{Phi2001} were the first ones to call them ultra compact dwarf galaxies (UCDs), a term which is widely in use for this type of objects at the present. \citet{Dri2003} reported that these objects are not only distinct from the globular clusters in the Milky Way (MWGCs) by their higher $V$-band ($L_{V}$) luminosity, but also by their larger radii and higher dynamical $V$-band mass-to-light ($M/L_{V}$) ratios.  At the same time, there is no gap in luminosity between globular clusters (GCs) and UCDs \citep{Mie2002,Mie2004}. \citet{Has2005} discovered in the Virgo cluster massive compact star clusters with similar properties like the ones in the Fornax cluster, but called them dwarf-globular transition objects (DGTOs). Like \citet{Dri2003}, they state that the dynamical $M/L_{V}$ ratios of some of the objects they discovered are significantly higher than the ones of the MWGCs. \citet{Mie2006} concluded from the H$\beta$ indices of UCDs in the Fornax cluster that they are most likely of intermediate age, while \citet{Evs2007} found the H$\beta$ indices of UCDs in the Virgo cluster most consistent with old ages. Their stellar population has evolved passively for a long time in any case, which makes UCDs similar to most GCs and elliptical galaxies in this respect.

Several formation scenarios that account for the physical properties of the UCDs have been proposed:
\begin{enumerate}
\item UCDs are the mergers of many massive young clusters that formed in a star burst triggered by a galaxy-galaxy encounter (e.g. like in the Antennae). After $\simeq 10 \, \rmn{Gyr}$ of dynamical (and stellar) evolution, such an object would resemble a UCD \citep{Kro1998,Fel2002}.
\item UCDs are the most luminous GCs \citep{Mie2002}.
\item UCDs are the central parts of nucleated galaxies that were disrupted by tidal forces as they moved in the gravitational field of a larger galaxy. Only the tightly bound cores survived until the present times \citep{Zin1988,Bas1994,Bek2003,Goe2007}.
\item UCDs are the remnants of the fundamental building blocks in galaxy formation \citep{Dri2004}.
\end{enumerate}

Some bright UCDs in the Fornax cluster and the Virgo cluster have been analysed by \citet{Hil2007} and \citet{Evs2007} very recently. They provide detailed high-quality data for 11 UCDs with dynamical masses between $10^7 \, \rmn{M}_{\odot}$ and $10^8 \, \rmn{M}_{\odot}$. Similar data have been obtained by \citet{Rej2007} for compact objects in Centaurus~A, but mostly with masses between $10^6 \, \rmn{M}_{\odot}$ and $10^7 \, \rmn{M}_{\odot}$. They enlarge a sample by \citet{Has2005} in the Virgo cluster by 20 objects in the same mass range. Taken together, these data allow us to analyse the change of the internal parameters of massive compact objects with mass or luminosity in more detail than \citet{Dri2003} or \citet{Has2005}. Furthermore, a comparison to other dynamically hot stellar systems (i.e. stellar systems whose stars are on randomised orbits) becomes possible, since samples with similar measured quantities are available as well. The quantities that are considered here include their $M/L_{V}$ ratio, $\Upsilon_V$, and their projected (effective) half-light radius, $r_{\rmn{e}}$, in dependency of their dynamical mass.

Especially the dynamical $M/L_{V}$ ratios of the UCDs has caught the attention of astronomers lately. \citet{Evs2007} find the UCDs in their sample to be consistent with predictions from simple stellar population (SSP) models within the errors. \citet{Hil2007} note a tendency of the SSP models to under-predict the $M/L_{V}$ ratios if a stellar population consistent with observations in the solar neighbourhood is assumed. \citet{Has2005} find that some of the stellar systems they discuss have $M/L_{V}$ ratios that imply extreme stellar populations in these objects. They suggest a presence of DM in these objects, provided that they are in dynamical equilibrium. This contradicts scenario (i), in which UCDs form DM free. Also if UCDs are nothing but very luminous GCs (scenario ii), they would be expected to be DM free, since GCs of usual size are. The simulations by \citet{Bek2003} on scenario (iii) predict DM free UCDs, since the DM halo of the progenitor galaxy of the UCD is found to be disrupted by the tidal interactions with the host galaxy of the UCD. This stands in contrast to the results from similar simulations by \citet{Goe2007}, who found that a UCD can still be DM dominated if it is the stripped nucleus of a nucleated galaxy. Scenario (iv) also suggests dark matter in UCDs. A detailed analysis of the $M/L_{V}$ ratios of the UCDs and their comparison to different SSP models may therefore give insights on their origin.

The stellar population of the UCDs obviously plays a decisive role for the $M/L_{V}$ ratio that has to be expected. The stellar population of each stellar system is determined, aside from an influence by stellar and dynamical evolution, by the stellar initial mass function (IMF), $\xi(m)$,
\begin{equation}
dN \propto \xi(m)\, dm,
\end{equation}
where $m$ is the stellar initial mass and $dN$ the number of stars in the mass interval $[m, m+dm]$. The IMF has to be distinguished from the present day stellar mass function (PDMF) which gives the number density of stars in dependency of stellar masses today. The IMF is a very useful concept, especially for a dynamically unevolved stellar system, because the number of stars that formed in the mass interval $[m,m+dm]$ is conserved with time on the whole domain of the IMF. As a consequence, the PDMF and IMF are very similar for stars still on the main sequence at the present time. It turns out in Section \ref{Trelax} that UCDs can indeed be considered as dynamically unevolved stellar systems due to their mass and extension and therefore long relaxation time.

In the past, there have been numerous efforts to infer the shape of the IMF from the PDMF as observed in resolvable stellar populations. There is common agreement that these observations are compatible with the IMF originally proposed by \citet{Sal1955} for field stars in the solar neighbourhood: $\xi (m) \propto m^{-\alpha}$ with $\alpha=2.35$ for $0.4 \, \rmn{M}_{\odot} \apprle m \apprle 10\, \rmn{M}_{\odot}$. Later observations indicated that $\alpha$ is constant up to the highest observed stellar masses (which are between $120 \, \rmn{M}_{\odot}$ and $200 \, \rmn{M}_{\odot}$, \citealt{Wei2004,Oey2005,Fig2005}), but gets smaller below  $0.5 \, \rmn{M}_{\odot}$ (\citealt{Kro2001} and references therein). The IMFs we consider for the stellar populations of the UCDs are guided by these results.

With masses between $10^7 \, \rmn{M}_{\odot}$ and $10^8 \, \rmn{M}_{\odot}$ and half-light radii mostly below $50 \, \rmn{pc}$, UCDs may have formed containing, within no more than some ten pc, between $10^5$ and $10^6$ O-stars or an order of magnitude more if the IMF was top-heavy. This is a scale of star formation beyond current theoretical reach, and it is therefore interesting to study the stellar content of these objects to probe the very extreme physics of their formation.

Let us stress the importance of \emph{dynamical} mass estimates for a meaningful discussion of the $M/L_{V}$ ratios. This puts a hard constraint on the UCDs that can be included in this discussion since it requires high-resolution spectroscopy of faint objects. However, a dynamical mass estimate is independent from the total luminosity of the stellar system. Instead, the mass estimate is based on the surface brightness profile and the width of the spectral lines as described in detail in \citet{Hil2007}. Dynamical mass estimates clearly rely on a number of assumptions that cannot be verified easily, but mass estimates for unresolved stellar populations based on stellar population models do so as well. The true advantage of the dynamical mass estimates for this work is that they allow an independent estimate for the $M/L_{V}$ ratio that can be compared to theoretical predictions from stellar population models.

This paper is organised as follows. In Section \ref{Data} a sample of different dynamically hot stellar systems, including UCDs, is introduced. Section \ref{Dependency} is dedicated to the dependencies of internal parameters of dynamically hot stellar systems on their mass. The $M/L_{V}$ ratio of UCDs, GCs and elliptical galaxies is compared to the predictions from simple stellar population models in Section \ref{SSP}. While doing this, we take the influence of their metallicity on their luminosity into account. Section \ref{Discussion} contains a discussion of the transition from GCs to UCDs. Furthermore the reliability of our results concerning the $M/L_{V}$ ratio of UCDs is addressed. We conclude with Section \ref{Conclusion}.

\section{The data}
\label{Data}
One of the tasks performed in this paper is to compare UCDs to other dynamically hot stellar systems. This requires a set of data which spans over many orders of magnitude in dynamical mass. A homogeneous data sample is unfortunately not available due to the diversity of the objects. We therefore collect data from different sources in the literature, where comparable parameters have been measured or where at least a correlation between the measured data to the ones that are to be compared is known. In the following, we specify the sources for our data and how we derived the quantities we use in this paper from them, if necessary.

\subsection{Massive Compact Objects}
\label{MCOs}
It is convenient in this paper to introduce massive compact objects (MCOs) as a collective term for all stellar systems in the sample discussed here that should neither be denominated as MWGCs nor as elliptical galaxies. This definition of MCOs includes a number of objects that are considered as UCDs in other works. The motivation for the introduction of this term lies in the fact that the sample of objects discussed here also includes a number of objects which in their entirety seem to mark a transition between GCs and UCDs. This will become apparent below. A clear distinction between GCs and UCDs is thereby problematic here.

We differentiate the MCOs by the way their dynamical masses were estimated:

For the 19 MCOs listed in Tab.~\ref{tabUCDdata}, the mass estimate included the fitting of a density profile to each one of them individually. These 19 objects are 12 MCOs from the Virgo cluster, five UCDs from the Fornax cluster as well as two objects from the Local Group: $\omega$~Cen in the Milky Way and G1 in Andromeda. We consider $\omega$~Cen as an MCO instead of an MWGC because of its spread in [Fe/H], which sets it apart from every other star cluster in the halo of the Galaxy (e.g. \citealt{Kay2006,Vil2007}) We refer to them as \textquotedblleft MCOs with mass distribution modelling\textquotedblright.

We also include 20 objects in Centaurus~A from \citet{Rej2007} for which measurements of the velocity dispersion and at least one colour index are available. Tab.~\ref{tabCendata} lists their properties. Their mass in $\rmn{M}_{\odot}$ is calculated by using a virial mass estimator given in \citet{Spi1987}:
\begin{equation}
M_{\sigma} \simeq 10 \, G^{-1} r_{\rmn{e}} \sigma ^2,
\label{eqVirMass1}
\end{equation}
where $r_{\rmn{e}}$ is the projected half-light radius\footnote{Actually, it is the half-mass radius that enters into eq.~(\ref{eqVirMass1}), but we assume that the mass density follows the luminosity density whenever necessary. This allows us to identify the half-mass radius with the half-light radius.} in pc and $\sigma$ is the global velocity dispersion in $\rmn{pc} \, \rmn{Myr}^{-1}$. $G$ is the gravitational constant, which is $0.0045 \, \rmn{pc}^3 \, \rmn{M}_{\odot}^{-1} \, \rmn{Myr}^{-2}$ We refer to them as \textquotedblleft MCOs with global mass estimate\textquotedblright.

\begin{table*}
\caption{Properties of MCOs with masses from mass distribution modelling. The contents of the columns are the following. Column 1: The name given to the MCO (the same as in the source papers), Column 2: The projected half-light radius of the MCO, Column 3: The global velocity dispersion of the MCO, Column 4: The central velocity dispersion of the MCO, Column 5: The absolute magnitude of the MCO in the $V$-band, Column 6: The dynamical mass of the MCO, Column 7: The $M/L_{V}$ ratio of the MCO, Column 8: References to the papers that are the basis for our data: 1: \citet{Evs2007}, 2: \citet{Has2005}, 3: \citet{Hil2007}, 4: \citet{Bau2003a}, 5: \citet{Ven2006}, 6: \citet{Har1996}. Some errors are marked with an asterisk; they have not been published so far.}
\label{tabUCDdata}
\vspace{2mm}
\centering
\begin{tabular}{lr@{}l@{\,}l@{\,}r@{}llllr@{}l@{\,}r@{}ll@{\,}ll}
\hline
&&&&&&\\[-10pt]
Name  & \multicolumn{5}{l}{$r_{\rmn{e}}$} & $\sigma$ & $\sigma_0$ & $M_V$  & \multicolumn{4}{l}{$M$}        & \multicolumn{2}{l}{$M/L_{V}$}             & Ref\\[2pt]
& \multicolumn{5}{l}{[pc]} & [$\rmn{km} \, \rmn{s}^{-1}$] &  [$\rmn{km} \, \rmn{s}^{-1}$]  &  [mag] & \multicolumn{4}{l}{[$10^6 \rmn{M}_{\odot}$]} & \multicolumn{2}{l}{[$\rmn{M}_{\odot} \, \rmn{L}_{\odot}^{-1}$]} &    \\[2pt]
\hline
&&&&&&\\[-10pt]
VUCD1        & 11. & 3  & $\pm$ & 0 & .7* & $32.2 \pm 2.4$ &$39.3\pm 2.0$ & $-12.26$         & 28 & $.0\, \pm$ & 5  & .0  & 4.0  &$ \pm \, 0.7$ & 1 \\
VUCD3        & 18. & 7  & $\pm$ & 1 & .8* & $35.8 \pm 1.5$ &$52.2\pm 2.5$ & $-12.58$         & 50 & $.0\, \pm$ & 7  & .0  & 5.4  &$ \pm \, 0.9$ & 1 \\
VUCD4        & 22. & 0  & $\pm$ & 2 & .7* & $21.3 \pm 2.0$ &$26.9\pm 2.3$ & $-12.30$         & 24 & $.0\, \pm$ & 6  & .0  & 3.4  &$ \pm \, 0.9$ & 1 \\
VUCD5        & 17. & 9  & $\pm$ & 0 & .8* & $26.4 \pm 1.6$ &$32.5\pm 2.3$ & $-12.32$         & 29 & $.0\, \pm$ & 4  & .0  & 3.9  &$ \pm \, 0.6$ & 1 \\
VUCD6        & 14. & 8  & $\pm$ & 3 & .1* & $22.3 \pm 1.8$ &$29.6\pm 2.2$ & $-12.10$         & 18 & $.0\, \pm$ & 5  & .0  & 2.9  &$ \pm \, 0.9$ & 1 \\
VUCD7        & 96. & 8  & $\pm$ & 20& *   & $27.2 \pm 4.6$ &$45.1\pm 1.5$ & $-13.44$         & 88 & $.0\, \pm$ & 21 & .0  & 4.3  &$ \pm \, 1.1$ & 1 \\
S417         & 14. & 36 & $\pm$ & 0 & .36 & $26.4 \pm 2.7$ &$31.7\pm 1.4$ & $-11.78\pm 0.16$ & 27 & $.0\, \pm$ & 5  & .0  & 6.6  &$ \pm \, 1.5$ &  1,2 \\
UCD1         & 22. & 4  & $\pm$ & 1 & .0  & $27.1 \pm 1.8$ &$41.3\pm 1.0$ & $-12.19$         & 32 & $.1\, \pm$ & 3  & .6  & 4.99 &$ \pm \, 0.60$ & 3,1 \\
UCD2         & 23. & 2  & $\pm$ & 1 & .0  & $21.6 \pm 1.8$ &$31.3\pm 0.6$ & $-12.27$         & 21 & $.8\, \pm$ & 3  & .1  & 3.15 &$ \pm \, 0.49$ & 3,1 \\
UCD3         & 89. & 9  & $\pm$ & 6 & .0  & $25.0 \pm 3.4$ &$29.3\pm 1.2$ & $-13.57$         & 94 & $.5\, \pm$ & 22 & .0  & 4.13 &$ \pm \, 0.98$ & 3,1 \\
UCD4         & 29. & 6  & $\pm$ & 2 & .0  & $22.8 \pm 3.1$ &$37.3\pm 0.6$ & $-12.45$         & 37 & $.3\, \pm$ & 8  & .6  & 4.57 &$ \pm \, 1.11$ & 3,1 \\
UCD5         & 30. & 0  & $\pm$ & 2 & .5  & $18.7 \pm 3.2$ &$28.7\pm 0.8$ & $-11.99$         & 18 & $.0\, \pm$ & 5  & .0  & 3.37 &$ \pm \, 0.85$ & 3,1 \\
S314         & 3.  & 23 & $\pm$ & 0 & .19 & \dots &$35.3\pm 1.4$ & $-10.91\pm 0.16$ & 5  & $.8\, \pm$ & 1  & .0  & 2.94 &$\pm \, 0.68$ & 2 \\
S490         & 3.  & 64 & $\pm$ & 0 & .36 & \dots &$42.5\pm 2.7$ & $-11.00\pm 0.16$ & 8  & $.7\, \pm$ & 2  & .1  & 4.06 &$\pm \, 1.15$ & 2 \\
S928         & 23. & 16 & $\pm$ & 1 & .37 & \dots &$22.4\pm 1.0$ & $-11.58\pm 0.16$ & 21 & $.3\, \pm$ & 2  & .9  & 6.06 &$\pm \, 1.23$ & 2 \\
S999         & 20. & 13 & $\pm$ & 0 & .98 & \dots &$25.6\pm 1.4$ & $-11.08\pm 0.16$ & 21 & $.6\, \pm$ & 2  & .9  & 9.36 &$\pm \, 1.87$ & 2 \\
H8005        & 28. & 69 & $\pm$ & 0 & .55 & \dots &$10.8\pm 2.3$ & $-10.83\pm 0.16$ & 5  & $.5\, \pm$ & 2  & .3  & 2.98 &$\pm \, 1.35$ & 2 \\
G1           & 8.  & 21 &       &   &     & \dots &$25.1\pm 1.7$ & $-10.94$         & 8  & $.2\, \pm$ & 0  & .85 & 4.10 &$\pm \, 0.42$ & 4 \\
$\omega$~Cen & 6.  & 70 & $\pm$ & 0 & .28 & $16.0$ &$19.0\pm 1.5$ & $-10.29$         & 2  & $.5\, \pm$ & 0  & .1  & 2.5  &$\pm \, 0.3$ & 5,6 \\ 
\hline
\end{tabular}
\end{table*}

\begin{table*}
\caption{Properties of the compact objects in Centaurus~A. Here the mass was calculated by using the same mass estimator for all objects, namely eq.~(\ref{eqVirMass1}). All data are from \citet{Rej2007}. The contents of the columns are the following. Column 1: The identification of the object (like in \citet{Rej2007}), Column 2: The effective (projected half light) radius of the MCO, Column 3: The global velocity dispersion, Column 4: The estimated (dynamical) mass, Column 5: The $M/L_V$ ratio.}
\label{tabCendata}
\vspace{2mm}
\centering
\begin{tabular}{lrrrr}
\hline
&&&&\\[-10pt]
Name  & \multicolumn{1}{l}{$r_{\rmn{e}}$}  & \multicolumn{1}{l}{$\sigma$} &     \multicolumn{1}{l}{$M_{\sigma}$}             & \multicolumn{1}{l}{$M / L_{V}$}   \\[2pt]
& \multicolumn{1}{l}{[pc]}   &  \multicolumn{1}{l}{[$\rmn{km} \, \rmn{s^{-1}}$]}    &  \multicolumn{1}{l}{[$10^6 \rmn{M}_{\odot}$]} & \multicolumn{1}{l}{[$\rmn{M}_{\odot} \, \rmn{L}_{\odot}^{-1}$]}  \\[2pt]
\hline
&&&&\\[-10pt]
HGHH92-C7 & $7.5 \pm 0.1$ & $21.6 ^ {+1.0} _ {-2.6}$ &  $7.8 ^ {+0.7} _ {-1.9}$ & $3.3 ^ {+0.8} _ {-1.1}$ \\[2pt]
HGHH92-C11 & $7.8 \pm 0.1$ & $19.6 ^ {+0.9} _ {-2.3}$ &  $6.7 ^ {+0.6} _ {-1.6}$ & $5.7 ^ {+1.4} _ {-1.9}$ \\[2pt]
HHH86-C15 & $5.3 \pm 0.7$ & $11.1 ^{+0.7} _ {-0.7}$ &  $1.5 ^ {+0.2} _ {-0.5}$ & $2.3 ^ {+0.6} _ {-0.9}$ \\[2pt]
HGHH92-C17 & $5.7 \pm 0.1$ & $20.9 ^ {+1.6} _ {-1.6}$ &  $5.8 ^ {+0.5} _ {-1.4}$ & $3.8 ^ {+0.9} _ {-1.3}$ \\[2pt]
HGHH92-C21 & $7.0 \pm 0.1$ & $19.3 ^ {+0.8} _ {-2.3}$  & $5.8 ^ {+0.5} _ {-1.4}$ & $4.8 ^ {+1.1} _ {-1.6}$ \\[2pt]
HGHH92-C22 & $3.8 \pm 0.1$ & $17.9 ^ {+0.1} _ {-0.1}$  & $2.8 ^ {+0.3} _ {-0.7}$ & $3.0 ^ {+0.7} _ {-1.0}$ \\[2pt]
HGHH92-C23 & $3.3 \pm 0.1$ & $31.3 ^ {+1.4} _ {-3.9}$  & $7.2 ^ {+0.7} _ {-1.8}$ & $1.8 ^ {+0.5} _ {-0.6}$ \\[2pt]
HGHH92-C29 & $6.9 \pm 0.1$ & $16.1 ^ {+0.8} _ {-0.8}$  & $4.1 ^ {+0.4} _ {-1.0}$ & $4.4 ^ {+1.0} _ {-1.4}$ \\[2pt]
HGHH92-C36 & $3.6 \pm 0.3$ & $15.7 ^ {+1.9} _ {-1.9}$  & $2.0 ^ {+0.3} _ {-0.6}$ & $2.6 ^ {+0.6} _ {-0.9}$ \\[2pt]
HGHH92-C37 & $2.9 \pm 0.3$ & $12.6 ^ {+0.8} _ {-0.8}$  & $1.1 ^ {+0.1} _ {-0.3}$ & $1.5 ^ {+0.4} _ {-0.6}$ \\[2pt]
HHH86-C38 & $2.8 \pm 0.2$ & $14.2 ^ {+1.1} _ {-1.1}$  & $1.3 ^ {+0.2} _ {-0.4}$ & $1.8 ^ {+0.4} _ {-0.6}$ \\[2pt]
HGHH92-C41 & $4.5 \pm 0.1$ & $11.5 ^ {+1.3} _ {-1.3}$  & $1.4 ^ {+0.1} _ {-0.3}$ & $2.2 ^ {+0.5} _ {-0.7}$ \\[2pt]
HGHH92-C44 & $5.7 \pm 0.1$ & $13.1 ^ {+1.0} _ {-1.0}$  & $2.3 ^ {+0.2} _ {-0.6}$ & $3.9 ^ {+0.9} _ {-1.3}$ \\[2pt]
HCH99-2 & $11.4 \pm 1.1$ & $14.1 ^ {+0.5} _ {-0.5}$ &  $5.3 ^ {+0.7} _ {-1.5}$ & $4.5 ^ {+1.2} _ {-1.6}$ \\[2pt]
HCH99-15 & $5.9 \pm 0.2$ & $21.3 ^ {+1.7} _ {-1.7}$ &  $6.2 ^ {+0.6} _ {-1.5}$ & $3.4 ^ {+0.8} _ {-1.1}$ \\[2pt]
HCH99-16 & $12.1 \pm 0.6$ & $9.5 ^ {+1.4} _ {-1.4}$ &  $2.5 ^ {+0.3} _ {-0.6}$ & $2.8 ^ {+0.7} _ {-0.9}$ \\[2pt]
HCH99-18 & $13.7 \pm 0.3$ & $21.2 ^ {+1.1} _ {-1.1}$ & $14.3 ^ {+1.3} _ {-3.5}$ & $4.7 ^ {+1.2} _ {-1.6}$ \\[2pt]
HCH99-21 & $7.1 \pm 2.7$ & $10.6 ^ {+2.3} _ {-2.3}$ &  $1.9 ^ {+0.7} _ {-1.0}$ & $1.7 ^ {+0.7} _ {-1.0}$ \\[2pt]
R223 & $2.6 \pm 0.3$ & $14.4 ^ {+1.5} _ {-1.5}$ &  $1.3 ^ {+0.2} _ {-0.4}$ & $2.3 ^ {+0.6} _ {-0.9}$ \\[2pt]
R261 & $1.9 \pm 0.4$ & $14.6 ^ {+0.7} _ {-0.7}$ &  $1.0 ^ {+0.2} _ {-0.3}$ & $1.1 ^ {+0.3} _ {-0.4}$ \\[2pt]
\hline
\end{tabular}
\end{table*}

\subsection{Globular clusters}
\label{GCs}
We compare the MCOs to the MWGCs for which \citet{Lau2005} calculated dynamical $M/L_{V}$ ratios (listed in their table~13). Their value for the effective half-mass radius and their estimate of the dynamical $M/L_{V}$ ratio in the $V$-band for the King model is used in this work. By using the absolute magnitude in the $V$-band given in \citet{Har1996}, the cluster mass can be calculated from its $M/L_{V}$ ratio.

It can hardly be expected that such a limited sample is representative for GCs in general. Nevertheless, this seems to be the case to some extent, as surveys of extragalactic GC systems show (e.g. \citealt{Lar2001,Cha2004} and \citealt{Jor2005} concerning the radii of GCs, and \citealt{Ric2003} and \citealt{Jor2007} concerning the absolute magnitudes of GCs, which indicate their masses if a constant $M/L$ ratio for them is assumed). It therefore seems possible to take the distribution of the radii and the masses of the MWGS as a rough representation of GCs in general. The advantage of the chosen sample is that, as for the MCOs, mass estimates from velocity dispersions are available for them.

\subsection{Early-type galaxies}
\label{Es}
We also compare the MCOs to more massive dynamically hot stellar systems by making use of some of the data published by \citet{Ben1992}, i.e. their values for the central velocity dispersion, $\sigma_0$, the projected half-light radius, $r_{\rmn{e}}$ and the absolute magnitude in the $B$-band of elliptical galaxies and bulges of early-type spiral galaxies in their sample. \citet{Ben1992} give a simple formula for estimating the King mass from $r_{\rmn{e}}$ and $\sigma_0$, which we use as well for the objects from their paper:
\begin{equation}
M_{\sigma 0}=5 \, G^{-1} r_{\rmn{e}} \sigma_0^2 ,
\label{eqVirMass2}
\end{equation}
with $r_{\rmn{e}}$ in pc, $\sigma_0$ in $\rmn{pc} \, \rmn{Myr}^{-1}$ and $G=0.0045 \, \rmn{pc}^3 \, \rmn{M}_{\odot}^{-1} \, \rmn{Myr}^{-2}$.

If these objects are to be compared to the MCOs, their $V$-band luminosities have to be estimated from their $B$-band luminosities, since for the MCOs luminosities in the $V$-band are measured. It is known that there is a correlation between the luminosity and the colour of elliptical galaxies. However, given the weakness of this dependency, we think that accounting for it (e.g. with the data on colour of the same galaxies from \citealt{Ben1993}) would probably not pay the effort. This becomes evident, if the uncertainties connected to the mass estimates from eq.~(\ref{eqVirMass2}) especially are considered (see Section \ref{NoteMass}). Therefore, adopting a uniform $B-V$ colour index of 0.9 seems a reasonable approximation for the purpose of this paper.

To enhance the sample, data on nucleated dwarf elliptical galaxies from \citet{Geh2003} are included.

Data on dwarf spheroidal galaxies (dSphs) are also included. They are taken from \citet{Met2007}, their table~2, because their data on dSphs are more up to date than the ones in \citet{Ben1992}. The half-light radii of the dSphs are not listed in that table, but are usually found in the references given there (with the exception of And II, for which the half-light radius is taken from the paper by \citealt{Con2006}).

\subsection{A note on different dynamical mass estimators}
\label{NoteMass}

\begin{figure}
\centering
\includegraphics[scale=0.80]{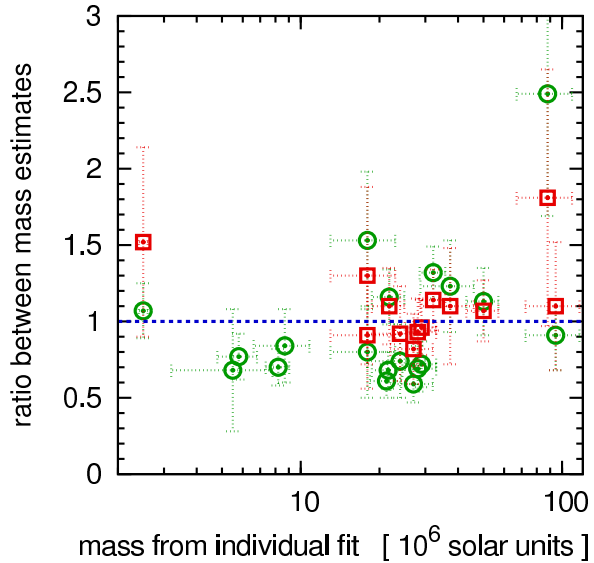}
\caption{Plot of the ratios between global estimates and mass estimates including mass distribution modelling for the 19 MCOs in Tab.~\ref{tabUCDdata} against the estimate for their mass from an individual fit. Open squares show $M_{\sigma}/M$ and circles show $M_{\sigma 0}/M$.}
\label{fig1}
\end{figure}

The dynamical mass of each of the objects introduced above was estimated in one of three different ways. While for some objects the mass estimate included the fitting of an individual density profile to them, for other objects the mass was calculated by using one of two global mass estimators. The choice of the mass estimator depended on whether $\sigma$ or $\sigma_0$ of the a stellar system was measured. This raises the question whether the mass estimates obtained in these different ways are indeed comparable. If they are comparable, two requirements should be fulfilled:
\begin{enumerate}
\item There should not be a tendency for one method to over- or underestimate the mass.
\item Applying different mass estimators on the same object should give similar results.
\end{enumerate}

This can be tested on the 19 MCOs in Tab.~\ref{tabUCDdata} where $\sigma$ and $\sigma_0$ or $\sigma_0$ only is available beside the mass estimate using an individual density profile, $M$, which is probably the most reliable one and therefore is considered as a standard here. Fig.~\ref{fig1} shows the masses as determined by using the global mass estimators in comparison to the mass from an individual density profile fit.

As a measure for the mean deviation of the mass estimated using eq.~(\ref{eqVirMass1}), $M_{\sigma}$, and the mass estimated using eq.~(\ref{eqVirMass2}), $M_{\sigma 0}$, from $M$, we calculate $\overline{\Delta M_{\sigma}} = \frac{1}{N_1} \, \sum_i^{N_1} |M_i-M_{\sigma \, i}|$ and $\overline{\Delta M_{\sigma 0}}= \frac{1}{N_2}\sum_i^{N_2} |M_i-M_{\sigma0 \, i}|$, where $N_1$ and $N_2$ denote the number of objects that are included for that summation. This results in $\overline{\Delta M_{\sigma}}=8.5 \times 10^6 \, M_{\odot}$ for the average deviation of $M_{\sigma}$ from $M$. This value can be compared to the mean value for the mass $\overline{M}$ of the same MCOs, with the masses as they are estimated using individual models for the density profile, which is $\overline{M} = 36.2 \times 10^6 \, M_{\odot}$. This means that the average deviation of $M_{\sigma}$ from $\overline{M}$ is about 23\%.

Similarly, the average deviation of $M_{\sigma 0}$ from $M$ can be calculated: $\overline{\Delta M_{\sigma 0}} = 12.5\times 10^6 M_{\odot}$. If this is again compared to $\overline{M}$ of the according MCOs, it turns out that the average deviation of $M_{\sigma 0}$ from $\overline{M}$ is about 44\%. The larger discrepancies between $M$ and $M_{\sigma 0}$ than between $M$ and $M_{\sigma}$ is at least partially due to the uncertainties to the inner density profiles of the MCOs, because the central structure of an MCO strongly influences the value that is determined for its $\sigma_0$.

The (relative and absolute) discrepancy between $M$ and $M_{\sigma}$ or $M_{\sigma 0}$ is the largest for VUCD7. However, VUCD7 is one of those MCOs that are best fit by a two-component (King+Sersic) density profile, in contrast to most of the other MCOs. It is therefore not surprising that the mass estimators eq.~(\ref{eqVirMass1}) and eq.~(\ref{eqVirMass2}) fail here, since they assume a King profile. This illustrates the risk connected to assuming a single typical profile for a number of objects. Excluding VUCD7, the average deviation of $M_{\sigma}$ from $\overline{M}$ can be lowered to about 10\%, and the average deviation of $M_{\sigma 0}$ from $\overline{M}$ can be lowered to about 24\%.

In summary, the three ways to estimate the dynamical mass seem to produce comparable results. Note that also \citet{Hil2007} and \citet{Evs2007} usually find that the internal parameters derived from global King estimators ($\alpha=2$) are almost identical to the parameters derived using mass distribution modelling. We will therefore not discriminate between $M_{\sigma}$, $M_{\sigma 0}$ and $M$ any further, but denote all dynamical masses as $M$.

\section{Dependencies on dynamical mass}
\label{Dependency}
In this section, the effective radii, median relaxation times, central densities and $M/L_{V}$ ratios are compared to each other.

\subsection{Dependency of the effective radius on mass}
\label{Radius}

\begin{figure*}
\centering
\includegraphics[scale=0.93]{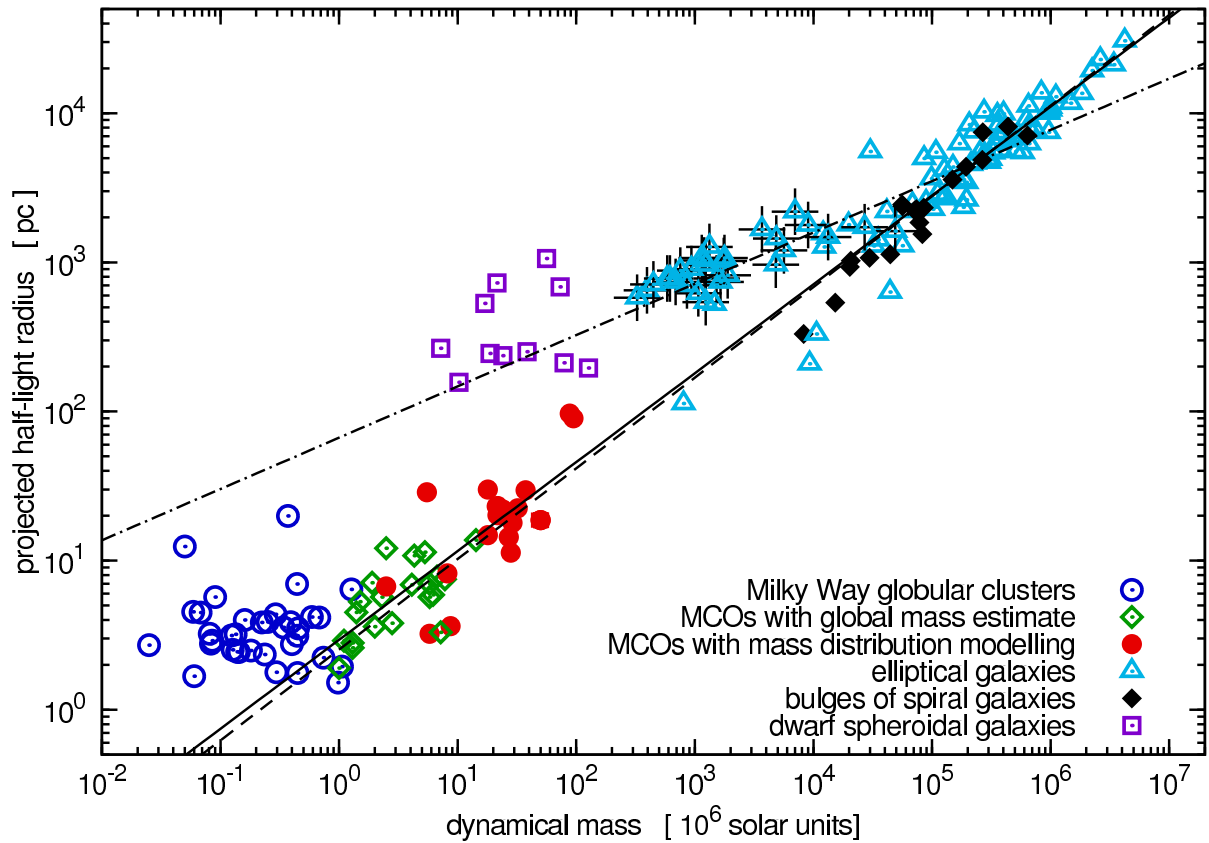}
\caption{Plot of the half-light radius, $r_{\rmn{e}}$, against mass, $M$, for different types of dynamically hot stellar systems. The symbols that are used have the following meaning: Open circles for MWGCs, open diamonds for MCOs with global mass estimate (i.e. calculated from eq.~\ref{eqVirMass1}), filled circles for MCOs with (the probably more reliable) mass estimates from mass distribution modelling (i.e. mass estimates taken from \citealt{Has2005}, \citealt{Hil2007} and \citealt{Evs2007} as well as the mass estimates for $\omega$ Cen and G1), open squares for dSphs, triangles for elliptical galaxies and filled diamonds for bulges of early-type spiral galaxies. Errors are comparable to the symbol sizes. The lines show fits to the data for a relation between mass and radius for bright ellipticals, compact ellipticals and bulges (dashed line), bright ellipticals, compact ellipticals, bulges and MCOs (solid line) and all elliptical galaxies, bulges and dSphs (dashed-dotted line). Most elliptical galaxies with low brightness have been excluded from the first two fits, see text for more details. They are marked with a cross. Note that the underlying assumption for the mass estimates is that the stellar systems are essentially undisturbed by tidal fields, which may be wrong for the dSphs especially \citep{Kro1997}.}
\label{fig2}
\end{figure*}

\begin{figure*}
\centering
\includegraphics[scale=0.93]{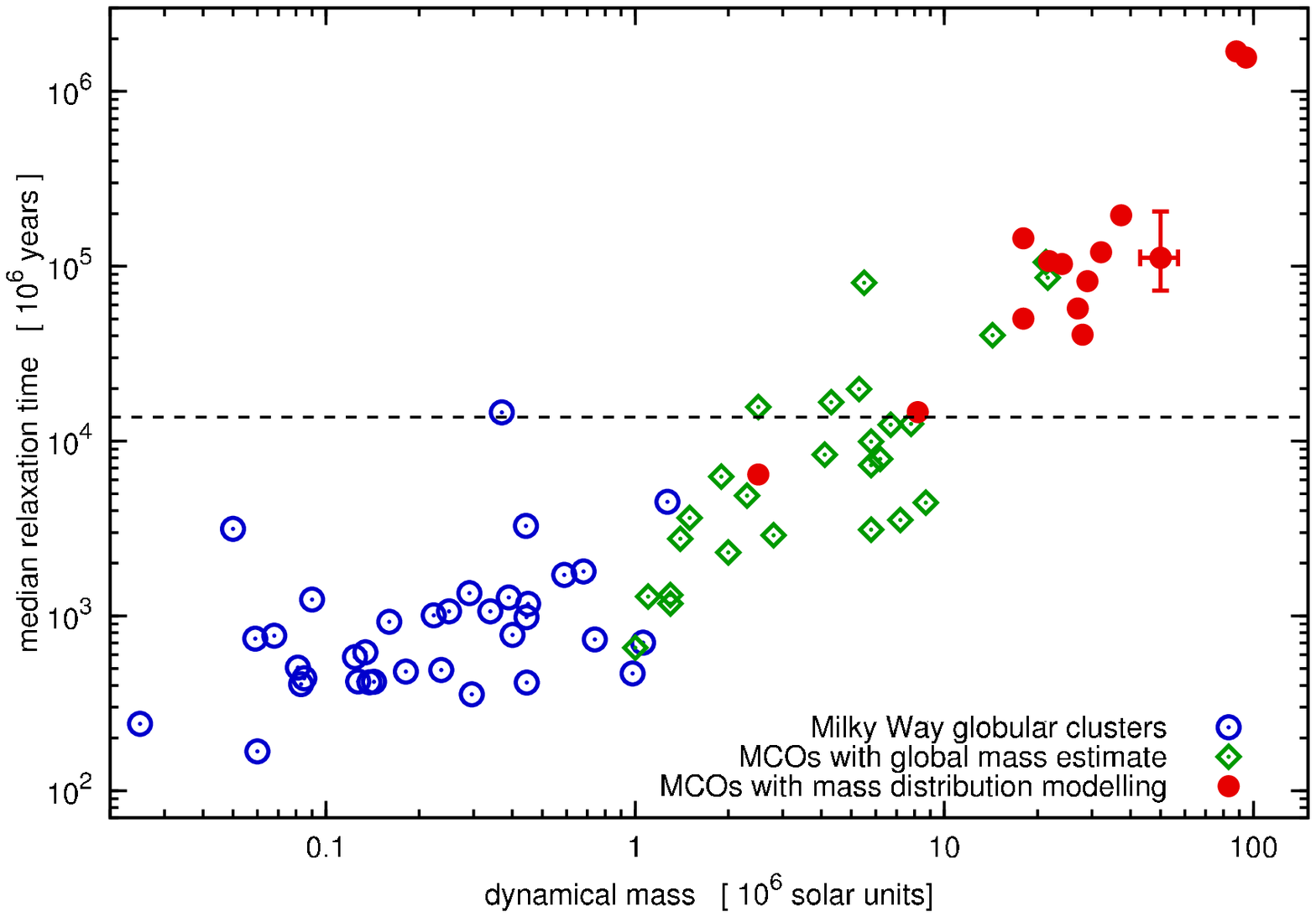}
\caption{The median relaxation time, $t_{\rmn{rel}}$, plotted against dynamical mass, $M$. Contrary to Fig. \ref{fig2}, this figure shows MWGCs and MCOs only. The dashed line marks the current age of the universe. The symbols are as in Fig.~\ref{fig2}. One MCO is plotted with typical errors.}
\label{fig3}
\end{figure*}

In Fig.~\ref{fig2}, the mass dependency of $r_{\rmn{e}}$ of the MCOs and other dynamically hot stellar systems is plotted. Some well established observations can be identified easily in this plot: The strong correlation between $M$ and $r_{\rmn{e}}$ for elliptical galaxies \citep{Ben1992} in the high mass range and the absence of a dependency of $r_{\rmn{e}}$ on $M$ for GCs \citep{Lau2000,Jor2005} at the lowest masses. Remarkable is the large spread of radii at intermediate masses which becomes largest in the mass interval of $10^7 \, \rmn{M}_{\odot} \apprle M \apprle 10^8 \, \rmn{M}_{\odot}$, the mass interval where the rather compact UCDs as well as the (typically about an order of magnitude) more extended dSphs lie. The underlying assumption for this statement is that dSphs are objects in (or close to) virial equilibrium. This has been argued to be the case by e.g. \citet{Wu2007} and \citet{Gil2007} for at least those dSphs that are most distant to the Galactic centre, although this would imply extremely high $M/L_{V}$ ratios in some cases. \citet{Gil2007} state that there is a bimodality of the characteristic radii of objects in the mass range $10^7 \, \rmn{M}_{\odot} \apprle M \apprle 10^8 \, \rmn{M}_{\odot}$, i.e. an almost complete absence of objects with $r_{\rmn{e}} \sim 100 \, \rmn{pc}$. In Fig.~\ref{fig2}, they are indeed only represented by VUCD7 and UCD3\footnote{Note that \citet{Gil2007} consider a half-light radius of only $22 \, \rmn{pc}$ (from \citet{Dri2003}) for UCD3 and omit VUCD7 from their discussion.} (and M32 at a higher mass). One way to interpret this is to consider UCDs and the dSphs as two kinds of stellar systems that formed under different conditions, as \citet{Gil2007} propose.

However, \citet{Met2007} argue that the formation of dSphs may have been triggered by the tidal forces in an encounter between two galaxies, i.e. they propose in principle the same scenario for the formation of dSphs which \citet{Fel2002} suggested for the formation of UCDs. The morphological differences can be understood in terms of the influence of the surroundings on the star-forming regions: the dSphs can form from star cluster complexes in a weak tidal field (e.g. the tidal arm of the Tadpole galaxy), while the UCDs form in a strong tidal field (e.g. the Antennae galaxy). This scenario is supported with the observation that the orbital angular momenta of the satellite galaxies of the Milky Way are correlated \citep{Met2008}. It can also offer an explanation for the seemingly high $M/L_{V}$ ratios of some of the dSphs, if they are largely unbound phase-space structures and therefore cannot be described by simple application of Jeans' equations \citep{Kro1997}.

It is surprising that the MCOs lie on the same relation between mass and radius as massive elliptical galaxies with masses $\apprge 10^{11} \, \rmn{M}_{\odot}$, while elliptical galaxies with lower masses (i.e. objects in the intermediate mass range) mostly lie on a different relation, which points towards the parameter space of dSphs. This could be evidence for the low-mass elliptical galaxies being mostly of tidal origin, as proposed by \citet{Oka2000} (also see fig. 7 in \citealt{Mon2007}), and as discussed by \citealt{Met2007} for dSphs. The few \emph{compact} low-mass elliptical galaxies can then be interpreted as low-mass counterparts of the elliptical galaxies more massive than $\apprge 10^{11} \, \rmn{M}_{\odot}$.

Following the above interpretation, some objects are thus excluded for quantifying the relation between mass and radius that MCOs share with massive elliptical galaxies in a least squares fit. These objects are, besides the MWGCs and the dSphs, the dwarf ellipticals from \citet{Geh2003} and the galaxies that \citet{Ben1992} define as \textquotedblleft bright dwarf ellipticals\textquotedblright\footnote{i.e. those galaxies which have $M_{V}>-18.5$ and are not classified as \textquotedblleft compact dwarf ellipticals\textquotedblright by \citet{Ben1992}}. The exclusion of the latter two groups may seem somewhat arbitrary, but it turns out that they define the apparent turn-off from the relation for the remaining objects (i.e. bright elliptical galaxies, galaxy bulges, compact ellipticals and MCOs) at $\apprge 10^{11} \, \rmn{M}_{\odot}$ quite well. Assuming a function of the form
\begin{equation}
\label{FunPlane}
\frac{r_{\rmn{e}}}{\rmn{pc}}=a\left( \frac{M}{10^6 \, \rmn{M}_{\odot}} \right)^b
\end{equation}
for the relation between $M$ and $r_{\rmn{e}}$, which corresponds to a straight line in Fig. \ref{fig2}, leads to
\begin{eqnarray}
\nonumber a&=&2.95^{+0.24}_{-0.22}, \\
\nonumber b&=&0.596 \pm 0.007,
\end{eqnarray}
for the best-fitting parameters. If the MCOs are not used for the fit,
\begin{eqnarray}
\nonumber  a&=&2.54^{+0.91}_{-0.67}, \\
\nonumber  b&=&0.608 \pm 0.025,
\end{eqnarray}
is obtained, i.e. within the errors the same relation as with the MCOs. The small impact that excluding the MCOs has on the fit is demonstrated in Fig. \ref{fig2} by plotting eq.~(\ref{FunPlane}) with both sets of values for $a$ and $b$. This veryfies that the MCOs lie along the same relation between $M$ and $r_{\rmn{e}}$ as massive elliptical galaxies.

For comparison, an analogous fit to \emph{all} elliptical galaxies as well as the dSphs (but without the MCOs) is performed. This corresponds to the hypothesis that these objects are drawn from a homogeneous population, which obeys a single relation between mass and radius. This leads to
\begin{eqnarray}
\nonumber  a&=&34.8^{+8.1}_{-6.6}, \\
\nonumber  b&=&0.399 \pm 0.019,
\end{eqnarray}
for the best-fitting parameters. However, the distribution of the massive elliptical galaxies is clearly asymmetric around this relation, which suggests that the first two relations are a better fit to them.

We note that the larger sample of elliptical galaxies which is used by \citet{Gra2006} shows a very similar distribution of characteristic radii against mass, although \citet{Gra2006} estimated the masses of the galaxies differently to the approach chosen here, namely by assuming a stellar population for them and calculating their masses from their luminosities.

The radii of MCOs are thus, unlike the ones of GCs, correlated to their masses. The comparison of the massive MCOs with the MWGCs shows that the characteristic radii of GCs are indeed typically about an order of magnitude smaller than the ones of the massive MCOs. However, Fig.~\ref{fig2} also seems to suggest a rather fluent transition between objects that lie on the scaling relation for GCs and objects that lie on the scaling relation for elliptical galaxies at a mass of about $10^6 \, \rmn{M}_{\odot}$. This confirms the conclusions \citet{Has2005} have drawn based on fewer data.

This change of typical radii cannot be due to an observational bias against small radii for more massive objects, since MCOs are identified by their brightness, their membership to a galaxy cluster and their \emph{compactness}. The data on rather low-mass MCOs from \citet{Has2005} and \citet{Rej2007} (both indicated as open diamonds in Fig.~\ref{fig2}) indeed include objects with radii on both scales. Consequently, this change of the typical $r_{\rmn{e}}$ must be connected to a difference in evolution or formation of objects less massive than $\approx 10^6 \, \rmn{M}_{\odot}$ and more massive than $\approx 10^7 \, \rmn{M}_{\odot}$.

\begin{figure*}
\centering
\includegraphics[scale=0.93]{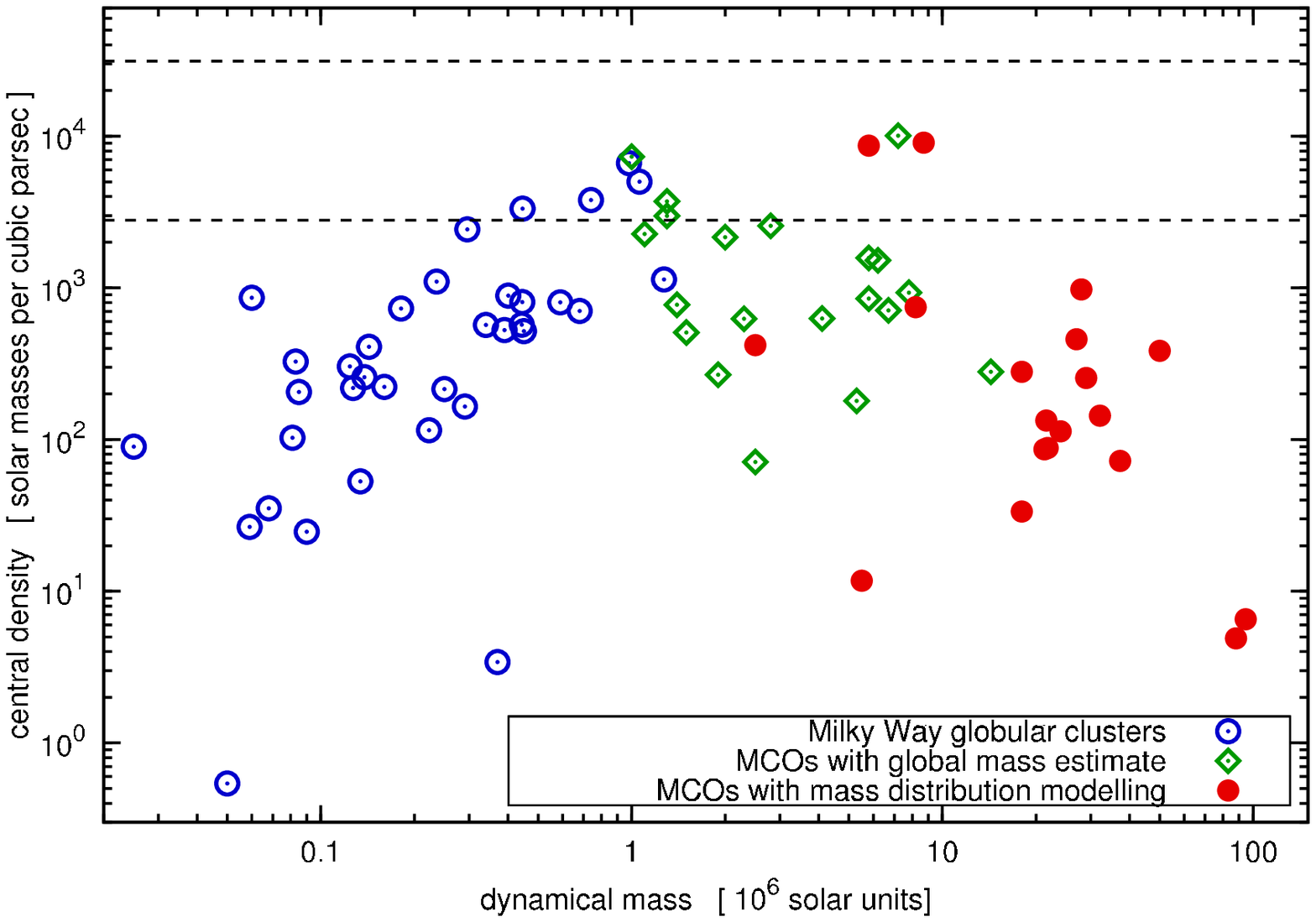}
\caption{The central density of MWGCs and MCOs plotted against dynamical mass. The symbols are as in Fig. \ref{fig3}. The dashed lines indicate constant densities: assuming a mean stellar mass of $0.4 \, \rmn{M}_{\odot}$, the lower dashed line indicates a density where the mean distance between stars is 6000 AU (about 100 times the diameter of the orbit of Neptune), and the upper dashed line indicates where the mean distance between stars is 3000 AU (about 50 times the diameter of the orbit of Neptune).}
\label{fig4}
\end{figure*}

\subsection{Dependency of the median two-body relaxation time on mass}
\label{Trelax}
The median two-body relaxation time is closely connected with mass and characteristic radius of an object. It is given in Myr in a formula originally found by \citet{Spi1971},
\begin{equation}
t_{\rmn{rel}}=\frac{0.061 \, N}{\log (0.4 \, N)} \times \sqrt{\frac{r_{\rmn{h}}^3}{GM}},
\label{eqRelax1}
\end{equation}
where $N$ is the number stars in the cluster, $r_{\rmn{h}}$ its half-mass radius in pc, $M$ its mass in $\rmn{M}_{\odot}$ and $G$ the gravitational constant, which is $0.0045 \, \rmn{pc}^3 \, \rmn{M}_{\odot}^{-1} \, \rmn{Myr}^{-2}$. $t_{\rmn{rel}}$ can be considered as a measure for the relaxation time at the half mass-radius. However, eq.~(\ref{eqRelax1}) is only an approximation: It is obtained under the assumption that the stars move in a smooth potential and are only disturbed by two-body encounters (i.e. no binaries), beside the supposition that the cluster is in virial equilibrium.

Eq.~(\ref{eqRelax1}) includes parameters which are not known for most of the MCOs, but it can be transformed into one that only depends on $M$ and the effective half-light radius, $r_{\rmn{e}}$, as free parameters if some assumptions are made. It can then be applied to the data in this paper. This is done by assuming that the mass is distributed as the luminosity and by substituting $r_{\rmn{e}}=0.75 \, r_{\rmn{h}}$ \citep{Spi1987}. We further assume a mean stellar mass of $0.4 \, \rmn{M}_{\odot}$ in concordance with the mean stellar mass in a stellar population with the canonical IMF (see eq.~\ref{canIMF}). This yields
\begin{equation}
t_{\rmn{rel}}=\frac{0.234}{\log (M/\rmn{M}_{\odot})} \times \sqrt{\frac{Mr_{\rmn{e}}^3}{G}}
\label{eqRelax2}
\end{equation}
in the same units as eq.~(\ref{eqRelax1}). An inspection of eq.~(\ref{eqRelax2}) reveals that $r_{\rmn{e}}$ dominates the behaviour of $t_{\rmn{rel}}$ due to its power. Therefore, a plot of $t_{\rmn{rel}}$ against $M$ looks very similar to a plot of $r_{\rmn{e}}$ against $M$ (Fig.~\ref{fig2}).

In Fig.~\ref{fig3}, $t_{\rmn{rel}}$ is plotted against $M$ of MWGCs and MCOs only. The stated similarity to Fig. \ref{fig2} in the according mass range is apparent. The new and important piece of information that can be read off Fig. \ref{fig3} is how $t_{\rmn{rel}}$ of the objects compares to a Hubble time. It is clearly below a Hubble time for most MWGCs, while it is clearly above a Hubble time for all MCOs more massive than $10^7 \, \rmn{M}_{\odot}$. This corresponds to the increase of the typical radii in the mass interval from $10^6 \, \rmn{M}_{\odot}$ to $10^7 \, \rmn{M}_{\odot}$. As MWGCs and MCOs are considered to be old objects, this implies that MWGCs can have undergone considerable dynamical evolution since their formation while massive MCOs have not. Consequently, massive MCOs are much less vulnerable to mass loss driven by two-body relaxation.

\subsection{Dependency of the central density on mass}
\label{Density}
It is worthwhile to consider the impact of the development of the typical radii with dynamical mass on  the central density of the MWGCs and MCOs. The central density is here defined as the mean density within the projected half-light (i.e. half-mass) radius. It is plotted in Fig. \ref{fig4} against mass.

The independence of the MWGC radii on their dynamical mass translates into an increase of the central density with dynamical mass. The increase of the typical radii above a dynamical mass of $10^6 \, \rmn{M}_{\odot}$, as visible in Fig. \ref{fig2}, is strong enough for a slow decrease of the central density to occur. It has already been noted by \citet{Bur1997} that there is a maximum global luminosity density for early-type galaxies, which is proportional to $M^{-4/3}$. In this light, the decrease of the densities with mass for the MCOs is only a consequence of the common relation between the MCOs and the massive elliptical galaxies that was found in Section \ref{Radius}.

\subsection{Dependency of the $M/L_{V}$ ratio on mass}
\label{MassLight}

\begin{figure*}
\centering
\includegraphics[scale=0.93]{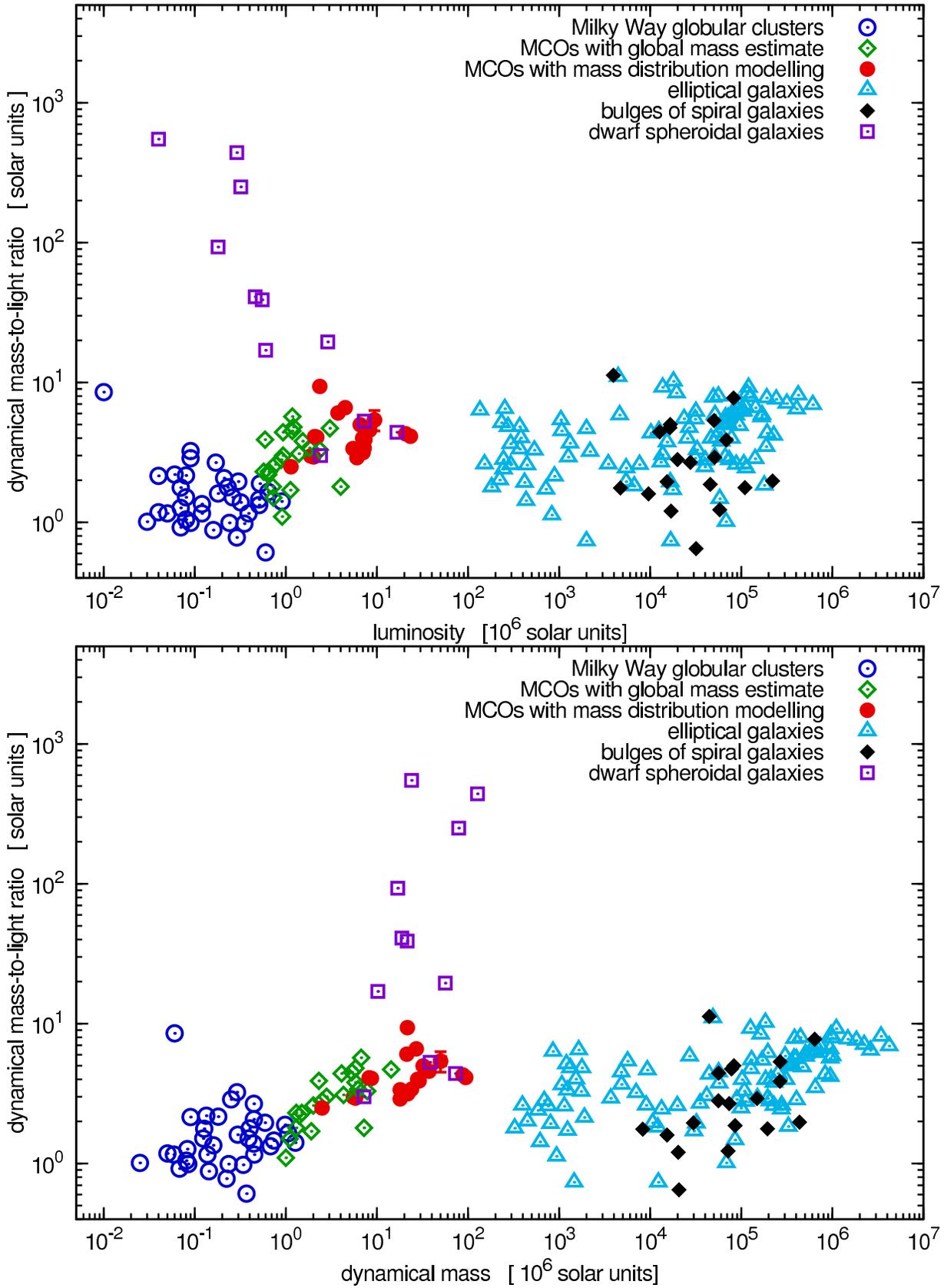}
\caption{Dynamical $M/L_{V}$ ratio plotted against luminosity in the $V$-band, $L_{V}$ (upper panel), and mass, $M$ (lower panel). The symbols are as in Fig.~\ref{fig2}. The errors to the values of the MCOs are not much larger than the symbol size.}
\label{fig5}
\end{figure*}

Fig. \ref{fig5} shows the dynamical $M/L_{V}$ ratios of the sample against the luminosity in the $V$-band (upper panel) and the dynamical mass (lower panel). It is visible from this figure that the dSphs with the lowest $V$-band luminosities also have the highest $M/L_{V}$ ratios, as was already noted in \citet{Mat1998}. Other than that, the general distribution of the data in both panels is almost identical, except for a steeper rise of the $M/L_{V}$ ratios from the MWGCs to the MCOs when they are plotted against $L_{V}$.

It might be tempting to identify the gap in the luminosity sequence at $\approx 10^8 \, \rmn{L}_{\odot}$ with the borderline between a star cluster-like population to the left and a galaxy-like population to the right. However, the \emph{homogeneous} sample of faint early-type galaxies in the Fornax cluster observed by \citet{Hil2003} does not show such a gap in luminosity down to the luminosities of dSphs. The gap visible in Fig. \ref{fig5} is thus most likely an artefact caused by the inhomogeneity of our data sample.

The spread of the $M/L_{V}$ ratios of the dSphs is very striking in Fig. \ref{fig5}. With $M/L_{V}$ ratios of several $10^2 \, \rmn{M}_{\odot}/\rmn{L}_{\odot}$, some of them are total outliers compared to all other dynamically hot stellar systems. It is especially the spread of their $M/L_V$ ratios that supports the notion that dSphs cannot be treated as objects in dynamical equilibrium. If they were in dynamical equilibrium, DM haloes with very different properties would have to be assumed for objects that are quite similar to each other as far as the properties of their baryonic matter are concerned.

It can be seen for the remaining objects that almost every MCO above a mass of $10^6 \, \rmn{M}_{\odot}$ has a $M/L_{V}$ ratio which is manifestly higher than the mean value for MWGCs. As for the radii, the transition from the $M/L_{V}$ ratios of GCs to the ones of MCOs seems fluent. The objects classified as some kind of elliptical galaxy (including bulges of early-type spiral galaxies) span the whole range of $M/L_{V}$ ratios that is occupied by GCs \emph{and} MCOs, with bulges and large elliptical galaxies having a larger spread to higher $M/L_{V}$ ratios.

It should be remembered in this context that the masses of the early-type galaxies that are used to determine their $M/L_V$-ratios have been calculated with eq.~(\ref{eqVirMass2}), i.e. the mass estimates are based on the distribution of the visible matter. If these galaxies are embedded in DM haloes, the mass estimates are too low for the total masses of the galaxies, but are still good approximations for the mass of those parts of the galaxies that are dominated by baryonic matter. It is noteworthy that evidence for DM only emerges in objects with $t_{\rmn{rel}}>\tau_{\rmn{H}}$ (also see Fig.~\ref{fig10} and \ref{fig12}).

The physical reasons for the distribution of $M/L_{V}$ ratios are a rather complicated issue. It mainly depends on two things: A possible non-baryonic DM content in the objects and the stellar populations of the objects. The $M/L_{V}$ ratio of a stellar population is influenced by its star formation history, its IMF, the metallicity of the stars and by how much the stellar population was altered by dynamical evolution. Unfortunately, most of the objects in our sample cannot be resolved into stars so far, which makes it impossible to determine their stellar populations directly. Nevertheless, observations of these objects and theoretical considerations can give some clues on their stellar populations. Some of these findings are summarised below.

\begin{itemize}
\item MWGCs contain old stellar populations (older than $\approx 10 \, \rmn{Gyr}$, \citealt{Van2000,Sal2002}). The MCOs in the Virgo cluster seem to have similar ages \citep{Evs2007}, but the MCOs in the Fornax cluster might be a bit younger \citep{Mie2006}. The ages of elliptical galaxies are found to range from a few Gyr to $\apprge 10 \, \rmn{Gyr}$ \citep{Tra2000,Ann2007}.
\item MWGCs are known to have low metallicities. The metallicities of the MCOs are, if estimated, consistent with those of metal-rich MWGCs. Elliptical galaxies have about solar metallicities in their central parts \citep{Tra2000,Ann2007} and a decrease of their metallicities towards their outer regions \citep{Tan1998,Bae2007}.
\item Dynamical evolution can lower the $M/L_{V}$ ratio of a stellar system noticeably, if the time scale for its dynamical evolution is shorter than the time scale for the evolution of its stars \citep{Bau2003b,Bor2007}.
\end{itemize}

With this information, the different $M/L_{V}$ ratios of the objects plotted in Fig.~\ref{fig5} become understandable at least qualitatively. The rather low $M/L_{V}$ ratio of MWGCs can be understood as an effect of their low metallicity and the considerable dynamical evolution that was suggested for them in section \ref{Trelax}. Considering the lifetimes \citet{Bau2003b} expect for a sample of MWGCs (while accounting for the tidal field of the Galaxy) and their results for the development of the $M/L$ ratio as a function of the star cluster lifetime, a decrease of the $M/L$ ratio by about $0.1 \, \rmn{M}_{\odot} \, \rmn{L}_{\odot}^{-1}$ to $0.3 \, \rmn{M}_{\odot} \, \rmn{L}_{\odot}^{-1}$ compared to the $M/L$ ratio of a dynamically unevolved stellar population would seem typical for MWGCs. The massive MCOs and the elliptical galaxies on the other hand are more metal-rich and due to their size and extension dynamically almost unevolved. This might be able to explain higher $M/L_{V}$ ratios compared to MWGCs even if they do not contain DM. Note however that a DM content in elliptical galaxies has been discussed: quite recently, \citet{Cap2006} estimated a median DM content of $\approx 30\%$ within the half-light radii of a sample of elliptical galaxies, if an IMF as in the Solar neighbourhood is assumed\footnote{\citet{Cap2006} do not discuss gas as a possible contributor to the non-luminous matter. However, considering the results by \citet{Com2007}, the mass of the gas is probably indeed negligible for their sample of galaxies. \citet{Com2007} estimate the mass of the molecular gas for the same sample of galaxies and find masses of the order of some $10^7 \, \rmn{M}_{\odot}$, which is about 3 to 4 orders of magnitudes less than the results in \citet{Cap2006} suggest for the total masses of the galaxies.}. The large spread of the $M/L_{V}$ ratios of ellipticals is not surprising in the light of their large age spread. Also recall the metallicity gradient in elliptical galaxies, which is natural if they are more complex than MWGCs and thus more diverse in their internal properties.

\section{The observed $M/L_{V}$ ratios and predictions of stellar population models}
\label{SSP}

For the remainder of this paper, we will compare the observed $M/L_{V}$ ratios of the objects discussed in the previous sections to predictions from stellar population models, with the focus on the $M/L_{V}$ ratios of the MCOs.

\subsection{The MCOs as simple stellar populations}
\label{Met}

In order to find which stellar population models are appropriate for the MCOs, we recall that most of the objects discussed here are old and note that a super-solar abundance of $\alpha$-elements seems to be typical for the dynamically hot stellar systems discussed here, see e.g \citet{Car1996} for MWGCs, \citet{Evs2007} for MCOs, and \citet{Ann2007} for elliptical galaxies. Therefore, self-enrichment through the ejecta of type I supernovae (SNI) apparently does not play a major role in these systems, as SNI are important contributors of iron to the interstellar medium \citep{Mat1986}. This can be taken as an indicator for a stellar population with a narrow age spread, if the progenitors of SNI are assumed to be white dwarfs that surpass the Chandrasekhar limit by accretion of additional matter \citep{Whe1973}. \citet{Mat2001} and \citet{Gre2005} suggest median time scales between some ten Myr and a few Gyr for the evolution of white dwarfs into SN I, depending on the initial conditions for the population. Considering stellar systems with ages of $\approx 10 \, \rmn{Gyr}$, this can be taken as a rather short time scale. The assumption of populations of coeval stars within each stellar system thereby seems a reasonable approximation for at least MWGCs and MCOs.

Besides age and age spread of the stars, a discussion of the $M/L_{V}$ ratios of stellar systems has to account for the metallicities of their stars, since the metallicity is known to have a influence on the colour and the luminosity of a star with a given mass. Therefore the metallicities of the stars have to be known if one intends to construct a model for a stellar population which accurately describes a real stellar population, including its $M/L_{V}$ ratio.

In the following, two assumptions for the metal abundances in the stellar populations of the MCOs are made. This is not only for the sake of simplicity but also for the lack of more detailed data in most cases.

Firstly, it is assumed that the metallicity-luminosity dependency of the stellar system can be characterised by the mean metallicity $Z$ of the stellar system. This would certainly be the case if $Z$ was equal to the metallicities of the component stars, i.e. if all stars had the same metallicity. However, this is not necessarily the case for the stars in MCOs, as the examples of $\omega$~Cen (e.g. \citealt{Kay2006,Vil2007}) and G1 \citep{Mey2001} show. On the other hand, imposing a more complicated metallicity distribution on the stars of the unresolved stellar populations of the other MCOs does not seem reasonable.

Secondly, it is assumed that the mean iron abundance, [Fe/H], allows solid conclusions on $Z$. This assumption can be motivated with the finding that $[\alpha/\rmn{Fe}] \simeq 0.3$ seems not only to be true for MWGCs \citep{Car1996}, but also for most of the MCOs that were analysed by \citet{Evs2007}. This value appears to be very typical for massive, dense star clusters.

The approximations and assumptions that have been made here and in Section \ref{MassLight} imply in their entirety that the stellar populations in MCOs can be considered as simple stellar populations (SSPs), meaning that all stars and stellar remnants have the same age and the same chemical composition.

\subsection{The metallicities of the MCOs}
\label{MetSam}

\begin{table}
\caption{MCOs with published metallicity estimates. [Fe/H] in Column 3 is taken as the measure for $Z$ of the object. Also $(V-I)$ colour indices are given for some objects whose metallicities were derived from line indices. They provide the opportunity to test the validity of eq.~(\ref{eqMet1}) on a sample of MCOs (see section \ref{Uncertainties}). The columns of the table contain the following information: Column 1: The name of the object, Column 2: [Z/H] if given in the reference, Column 3: [Fe/H] either from the reference or calculated using eq.~(\ref{eqFeAlpha}), Column 4: The $(V-I)$ colour index, Column 5: The reference to the source paper: 1: \citet{Evs2007}, 2: \citet{Mie2006}, 3: \citet{Has2005}, 4: \citet{Mey2001}, 5: \citet{Har1996}.}
\label{tabUCDmet}
\vspace{2mm}
\begin{tabular}{@{\,}lr@{\,}c@{\,}rr@{}l@{\,}lll}
\hline
&&&\\[-10pt]
Name & \multicolumn{3}{l}{[Z/H]} & \multicolumn{3}{l}{[Fe/H]} & $(V-I)$ & Ref.\\ [2pt]
\hline
&&&\\[-10pt]
VUCD1        & $-1.35$ &\dots &$-0.33$ & $-1.$ & 12  & $\pm \, 0.51$   & 0.96 & 1\\
VUCD3        &  0.00   &\dots &0.35    & $-0.$ & 107 & $\pm \, 0.175$  & 1.27 & 1\\
VUCD4        & $-1.35$ &\dots &0.33    & $-1.$ & 12  & $\pm \, 0.51$   & 0.99 & 1\\
VUCD5        & $-0.33$ &\dots &0.00    & $-0.$ & 447 & $\pm \, 0.165$  & 1.11 & 1\\
VUCD6        & $-1.35$ &\dots &$-0.33$ & $-1.$ & 12  & $\pm \, 0.51$   & 1.02 & 1\\
VUCD7        & $-1.35$ &\dots &$-0.33$ & $-1.$ & 12  & $\pm \, 0.51$   & 1.13 & 1\\
S417         & $-1.35$ &\dots &0.00    & $-0.$ & 957 & $\pm \, 0.65$   && 1\\
UCD1         &         &      &        & $-0.$ & 38  & $\pm \, 0.05$   & 1.11 & 2\\
UCD2         &         &      &        & $-0.$ & 90  & $\pm \, 0.33$   & 1.12 & 2\\
UCD3         &         &      &        & $-0.$ & 52  & $\pm \, 0.11$   & 1.18 & 2\\
UCD4         &         &      &        & $-0.$ & 85  & $\pm \, 0.29$   & 1.12 & 2\\
UCD5         &         &      &        & \multicolumn{2}{c}{\dots}& && \\
S314         &         &      &        & $-0.$ & 50  &              && 3\\
S490         &         &      &        &  0.   & 18  &              && 3\\
S928         &         &      &        & $-1.$ & 34  &              && 3\\
S999         &         &      &        & $-1.$ & 38  &              && 3\\
H8005        &         &      &        & $-1.$ & 27  &              && 3\\
G1           &         &      &        & $-0.$ & 95  & $\pm \, 0.09$ && 4\\
$\omega$ Cen &         &      &        & $-1.$ & 62  &              && 5\\
\hline
\end{tabular}
\end{table}

\begin{table*}
\caption{Colours and derived [Fe/H] for the Centaurus~A objects. The contents of the columns in the table are the following: Column 1: Identification of the object like in \citet{Rej2007}, Column 2: The $(V-I)$ colour index, Column 3: The $(B-V)$ colour index, Column 4: [Fe/H] calculated from the $(V-I)$ colour index, Column 5: [Fe/H] calculated from the $(B-V)$ colour index, Column 6: Our final estimate for [Fe/H] with the adopted errors.}
\label{tabCenmet}
\vspace{2mm}
\centering
\begin{tabular}{llllll}
\hline
&&&&&\\[-10pt]
Name & ($B-V$) & ($V-I$) & $\rmn{[Fe/H]}_{(B-V)}$ & $\rmn{[Fe/H]}_{(V-I)}$ & [Fe/H] \\ [2pt]
\hline
&&&&&\\[-10pt]
HGHH92-C7 & 0.75 & 0.91 & $-1.13$ & $-1.55$ & $-1.34 \pm 0.30$ \\
HGHH92-C11 & 0.94 & 1.12 & $-0.09$ & $-0.66$ & $-0.38 \pm 0.39$ \\
HHH86-C15 & 0.89 & 1.03 & $-0.36$ & $-1.04$ & $-0.70 \pm 0.42$ \\
HGHH92-C17 & 0.77 & 0.88 & $-1.02$ & $-1.68$ & $-1.35 \pm 0.39$ \\
HGHH92-C21 & 0.78 & 0.93 & $-0.97$ & $-1.47$ & $-1.22 \pm 0.33$ \\
HGHH92-C22 & 0.79 & 0.91 & $-0.91$ & $-1.55$ & $-1.23 \pm 0.39$ \\
HGHH92-C23 & 0.76 & 0.78 & $-1.08$ & $-2.10$ & $-1.59 \pm 0.55$ \\
HGHH92-C29 & 0.89 & 1.08 & $-0.36$ & $-0.83$ & $-0.60 \pm 0.35$ \\
HGHH92-C36 & 0.73 & 0.85 & $-1.24$ & $-1.80$ & $-1.52 \pm 0.35$ \\
HGHH92-C37 & 0.84 & 0.99 & $-0.64$ & $-1.21$ & $-0.93 \pm 0.37$ \\
HHH86-C38 & 0.78 & 0.91 & $-0.97$ & $-1.55$ & $-1.26 \pm 0.36$ \\
HGHH92-C41 & 0.89 & 1.09 & $-0.36$ & $-0.79$ & $-0.58 \pm 0.33$ \\
HGHH92-C44 & 0.69 & 0.85 & $-1.47$ & $-1.80$ & $-1.63 \pm 0.26$ \\
HCH99-2 & 0.74 & 0.84 & $-1.19$ & $-1.85$ & $-1.52 \pm 0.39$ \\
HCH99-15 & \dots & 1.06 & \dots & $-0.92$ & $-0.62 \pm 0.23$ \\
HCH99-16 & \dots & 0.79 & \dots & $-2.06$ & $-1.76 \pm 0.23$ \\
HCH99-18 & 0.89 & 0.89 & $-0.36$ & $-1.63$ & $-1.00 \pm 0.67$ \\
HCH99-21 & \dots & 0.78 & \dots & $-2.10$ & $-1.80 \pm 0.23$ \\
R223 & 0.80 & 0.95 & $-0.86$ & $-1.38$ & $-1.12 \pm 0.35$ \\
R261 & 0.83 & 0.99 & $-0.70$ & $-1.21$ & $-0.95 \pm 0.35$ \\
\hline
\end{tabular}
\end{table*}

Information on the metallicities of MCOs are published in \citet{Has2005}, \citet{Mie2006}, and \citet{Evs2007}. \citet{Evs2007} give for each of the MCOs they examined an interval in which the actual mean metallicity $Z$ of the MCO lies. We assume that this true value for $Z$ of the MCO lies in the middle of the interval given. \citet{Has2005} and \citet{Mie2006} do not give estimates for $Z$ of the objects they discuss, but for [Fe/H]. Based on the observational findings by \citet{Car1996} and \citet{Evs2007} and the assumption that the iron abundance characterises the metallicity of the MCOs, we adopt $[\alpha/\rmn{Fe}]=0.3$ for each one of them and use the relation
\begin{equation}
[Z/\rmn{H}]=\rmn{[Fe/H]}+0.94 \, [ \alpha/\rmn{Fe}]
\label{eqFeAlpha}
\end{equation}
found by \citet{Tho2003} to calculate $[Z/\rmn{H}]$ from [Fe/H].
The values that are adopted for the element abundances of the MCOs are summarised in Tab.~\ref{tabUCDmet}.

For the objects in Centaurus~A no metallicities have been published so far, but $(B-V)$ and $(V-I)$ colour indices for them are available in \citet{Rej2007}. Observations show that there is a correlation between colour indices and [Fe/H] in GC systems. On this basis, an estimate of [Fe/H] in the objects in Centaurus~A can be made by assuming that they follow a relation between colour and metallicity that has been established on another GC system. \citet{Bar2000} give relations between [Fe/H] and $(V-I)$ as well as [Fe/H] and $(B-V)$ for the GC system of the Milky Way, using the data from \citet{Har1996}:
\begin{equation}
\rmn{[Fe/H]}_{(V-I)} = (4.22 \pm 0.39)\times (V-I)-(5.39 \pm 0.35)
\label{eqMet1}
\end{equation}
and
\begin{equation}
\rmn{[Fe/H]}_{(B-V)} = (5.50 \pm 0.33)\times (B-V)-(5.26 \pm 0.23).
\label{eqMet2}
\end{equation}
The confidence range of these equations is set by the values $(V-I)$ and [Fe/H] can assume for MWGCs. Their values for [Fe/H] are mostly between $-2$ and $-0.5 \, \rmn{dex}$.

The advantage of the relations from \citet{Bar2000} is that they have been established for both colour indices that have been measured for the objects in Centaurus~A, i.e. they allow us to fully benefit from the available data. Their disadvantage is that they do not account for a slight curvature in the relation between [Fe/H] and the colour indices, which is typical for this relation according to \citet{Yoo2006}. However, given the apparent weakness of this departure from linearity, it seems justified to neglect it.

\begin{figure}
\centering
\includegraphics[scale=0.80]{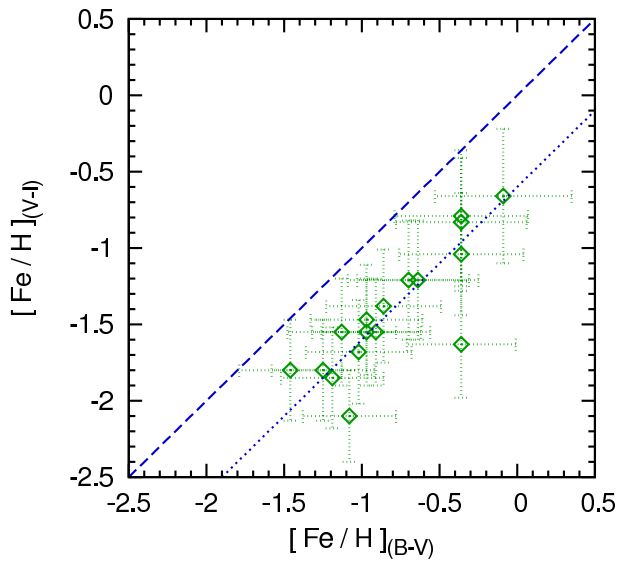}
\caption{Comparison between $\rmn{[Fe/H]}_{(V-I)}$ and $\rmn{[Fe/H]}_{(B-V)}$ for the objects in Centaurus~A. The numbers have been calculated with eq.~(\ref{eqMet1}) and eq.~(\ref{eqMet2}) respectively. The dashed line indicates equality of $\rmn{[Fe/H]}_{(V-I)}$ and $\rmn{[Fe/H]}_{(B-V)}$. The dotted line, corresponding to $\rmn{[Fe/H]}_{(V-I)}=\rmn{[Fe/H]}_{(B-V)}-0.6$, is a fit by eye to the actual distribution of the data.}
\label{fig6}
\end{figure}

We calculate $\rmn{[Fe/H]}_{(V-I)}$ and $\rmn{[Fe/H]}_{(B-V)}$ for each cluster in Centaurus~A from eq.~(\ref{eqMet1}) and (\ref{eqMet2}) if both colour indices are available. The results from eq.~(\ref{eqMet1}) turn out to be systematically lower by $\approx 0.6 \, \rmn{dex}$ on average than the results from eq.~(\ref{eqMet2}), as can be seen in Fig.~\ref{fig6}. It is obvious that the different results for the iron abundance calculated from different colour indices may indicate a serious problem with those estimates. A discussion on how reliable the results based on these metallicity estimates are will be given Section \ref{Discussion}. For now, we clearly distinguish between objects with [Fe/H] estimates from colour indices and objects with [Fe/H] estimates from line indices.

The relation between [Fe/H] and $(V-I)$ colour found by \citet{Kis1998} by including (beside MWGCs) GCs around NGC 1399 has a slightly flatter slope than eq.~(\ref{eqMet1}). It yields however similar results in the colour range interesting for the purpose here (deviations would be $\approx 0.2 \, \rmn{dex}$ in the most extreme cases).

As a compromise between the two values that are estimated for the iron abundances of the objects in Centaurus~A, we adopt the mean of both values as our final value for [Fe/H]. The error to this value has two components. The first of them is due to the intrinsic uncertainties to eqs.~(\ref{eqMet1}) and (\ref{eqMet2}). The second component is the uncertainty due to the systematic difference between the results from eqs.~(\ref{eqMet1}) and (\ref{eqMet2}). We estimate this error as half the difference between both estimates for a particular object. For the total error to the estimate of [Fe/H], the square root of the sum of the squares of both errors is assumed.

For three objects only a $(V-I)$ colour index is given. In these cases we simply set $\rmn{[Fe/H]}_{(V-I)}+0.3 \, \rmn{dex}=\rmn{[Fe/H]}$, as $0.3 \, \rmn{dex}$ is the average value by which the $(V-I)$ colour indices of the other objects are changed. For estimating an error to these values for [Fe/H], the scatter of the data for $\rmn{[Fe/H]}_{(V-I)}$ and $\rmn{[Fe/H]}_{(B-V)}$ around the relation $\rmn{[Fe/H]}_{(V-I)}=\rmn{[Fe/H]}_{(B-V)}-0.6$ is calculated for the objects in Fig~\ref{fig6}. The scatter, $s$, is given by the equation $s^2 = \frac{1}{N-1} \, \sum_i^{N} [\rmn{[Fe/H]}_{(V-I) \, i}-(\rmn{[Fe/H]}_{(B-V) \, i}-0.6 \, \rmn{dex})]^2$, where $N=17$ is the number of objects in Fig.~6. This results in $s=0.23 \, \rmn{dex}$, which we adopt as the error to the [Fe/H] values of these three objects.

The numbers for the metallicities of the objects in Centaurus~A are listed in Tab.~\ref{tabCenmet}.

\begin{figure}
\centering
\includegraphics[scale=0.80]{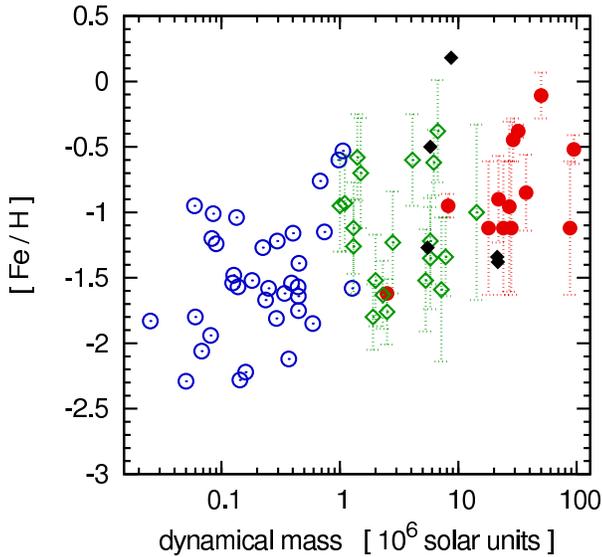}
\caption{The iron abundances adopted for this work plotted against the dynamical mass for the MWGCs and the MCOs. Open circles represent the MWGCs, filled circles the MCOs with abundance estimates from line indices, open diamonds the MCOs in Centaurus~A \citep{Rej2007} and filled diamonds the MCOs in the Virgo cluster from \citet{Has2005}. The values for [Fe/H] of the latter two have been calculated from colour indices.}
\label{fig7}
\end{figure}

Note that \citet{Has2005} obtain the [Fe/H] estimates for their objects also by comparison of the colour indices to the ones of GCs, i.e. in very much the same fashion as is done here for the objects in Centaurus~A. The only MCOs with abundance estimates from line indices and thus estimates directly linked to an actual presence of the according elements in the cluster are the objects from \citet{Evs2007}, the objects from \citet{Hil2007}, $\omega$ Cen and G1.

Since Figs.~\ref{fig2},~\ref{fig3},~\ref{fig4}~and~\ref{fig5} suggest a rather fluent transition from the properties of MWGCs to the ones of MCOs, it seems worthwhile to include them in the discussion further on. A comprehensive compilation of the iron abundances of MWGCs is provided by \citet{Har1996}. Based on the results of \citet{Car1996}, we assume $[\alpha/\rmn{Fe}]=0.3$ in order to calculate $Z$ for them, as we did for the MCOs (eq.~\ref{eqFeAlpha}).\\
Like the ones of MCOs, the stellar populations of MWGCs can be considered as old and coeval, but, in contrast to the ones of MCOs, dynamically evolved (i.e. loss of low-mass stars though evaporation driven by two-body relaxation).

The [Fe/H] that are adopted for the MWGCs and the MCOs are plotted in Fig.~\ref{fig7}. A tendency to higher abundances with higher masses is undeniable. Note however that selection effects might play a role here. There is a bias against metal-rich objects for MWGCs, because they are concentrated towards the bulge of the Galaxy and therefore harder to observe than the metal-poor halo MWGCs \citep{Har1976}. The GC systems of elliptical galaxies, on the other hand, have a larger fraction of red (probably metal-rich) GCs, which are also somewhat brighter than the blue (probably metal-poor) ones \citep{Har2006,Weh2007}.

\subsection{Predictions for $M/L_{V}$ ratios from SSP models}
\label{CompMet}
If information on the dependency of $M/L_V$ ratio of a SSP on $Z$ is combined with the estimates on the metallicity of the MCOs, it can be appraised what differences in $\Upsilon_V$ are \emph{not} due to differences in $Z$. Theoretical estimates of $\Upsilon_V$ for different $Z$ are taken from \citet{Mar2005} for SSPs that formed with a canonical IMF or a Salpeter-Massey IMF and from \citet{Bru2003} for SSPs that formed with a Chabrier IMF.

\begin{figure}
\centering
\includegraphics[scale=0.76]{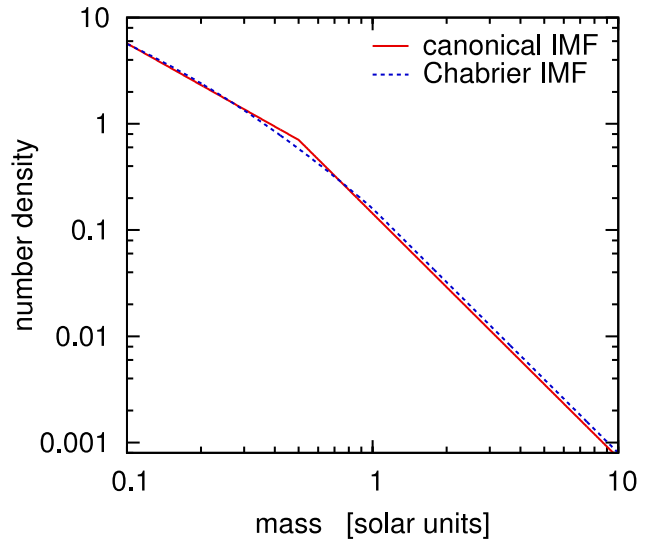}
\caption{A comparison between the canonical IMF and the Chabrier IMF in the interval from $0.1 \, \rmn{M}_{\odot}$ to $10 \, \rmn{M}_{\odot}$. Both IMFs are normalised such that $\int^{100}_{0.1} \xi(m)m \, dm=1$, where $m$ is the mass in solar units. The two IMFs are barely distinguishable on the whole mass interval. They actually are identical above a mass of $1 \, \rmn{M}_{\odot}$, except for a slightly different normalisation factor.}
\label{fig8}
\end{figure}

The canonical IMF is a continuous multi-power law,
\begin{equation}
\xi_{\rmn{K}}(m) \propto m^{-\alpha_i},
\label{canIMF}
\end{equation}
with $\alpha_1 =1.3$ for $m < 0.5 \, \rmn{M}_{\odot}$ and $\alpha_2= 2.3$ for $m > 0.5 \, \rmn{M}_{\odot}$. It has been constrained after a decade-long study of various biases and found to be consistent with all resolved stellar populations so far \citep{Kro1993,Kro2001,Kro2002,Kro2007}. The Chabrier IMF is given for $m < 1 \, \rmn{M}_{\odot}$ as
\begin{equation}
\xi_{\rmn{C}}(m) \propto \frac{1}{m} \, \exp \left[ - \frac{(\log (m/ \rmn{M}_{\odot})- \log 0.08)^2}{0.9522}\right]
\end{equation}
and equals the canonical IMF for $m > 1 \, \rmn{M}_{\odot}$ up to a normalisation factor. The transition at $1 \, \rmn{M}_{\odot}$ is continuous \citep{Cha2001,Cha2003}. This IMF cannot be distinguished from the canonical IMF within the observational errors (Fig.~\ref{fig8}). To simplify matters, we will therefore also refer to the Chabrier IMF as the canonical IMF. The Salpeter-Massey IMF is a single power law with $\alpha=2.35$ \citep{Sal1955,Mas1998}. The SSP models used here have been obtained under the assumption that the IMFs are defined from $0.1 \, \rmn{M}_{\odot}$ to $100 \, \rmn{M}_{\odot}$.

Note that the upper mass limit of the IMF as suggested by \citet{Wei2004}, \citet{Oey2005} and \citet{Fig2005} is higher than $100 \, \rmn{M}_{\odot}$, but this does not have a mentionable affect on the expected $M/L$ ratios of the SSPs discussed here due to the scarcity of high-mass stars in them.

A lower mass limit of $0.1 \, \rmn{M}_{\odot}$ for the IMF neglects the existence of brown dwarfs. This is probably unproblematic, if one follows the argumentation by \citet{Thi2007}. They suggest that star-like objects and brown dwarf-like objects are different populations and thus their frequencies cannot be described by a single, continuous IMF as e.g. in \citet{Kro2001}. The combined mass functions for brown dwarfs and stars which they find for star clusters in the Milky Way have many fewer brown dwarfs. Assuming a similar situation in the MCOs, brown dwarfs are not expected to contribute more than a few percent to their total mass (opposed to $\approx 10 \%$ for a mass function as in \citealt{Kro2001}).

The ages that are considered here for the SSPs are $9 \, \rmn{Gyr}$ and $13 \, \rmn{Gyr}$. Note that \citet{Mar2005} distinguishes between different horizontal branch morphologies, but this has a negligible impact on the dependency of $\Upsilon_V$ on $Z$ of an old SSP.

The benefit from using both the SSP models from \citet{Bru2003} and \citet{Mar2005} although they cover the same ages and use (in principle) the same IMF is that different stellar evolutionary models have been used for constructing them.

\begin{figure}
\centering
\includegraphics[scale=0.85]{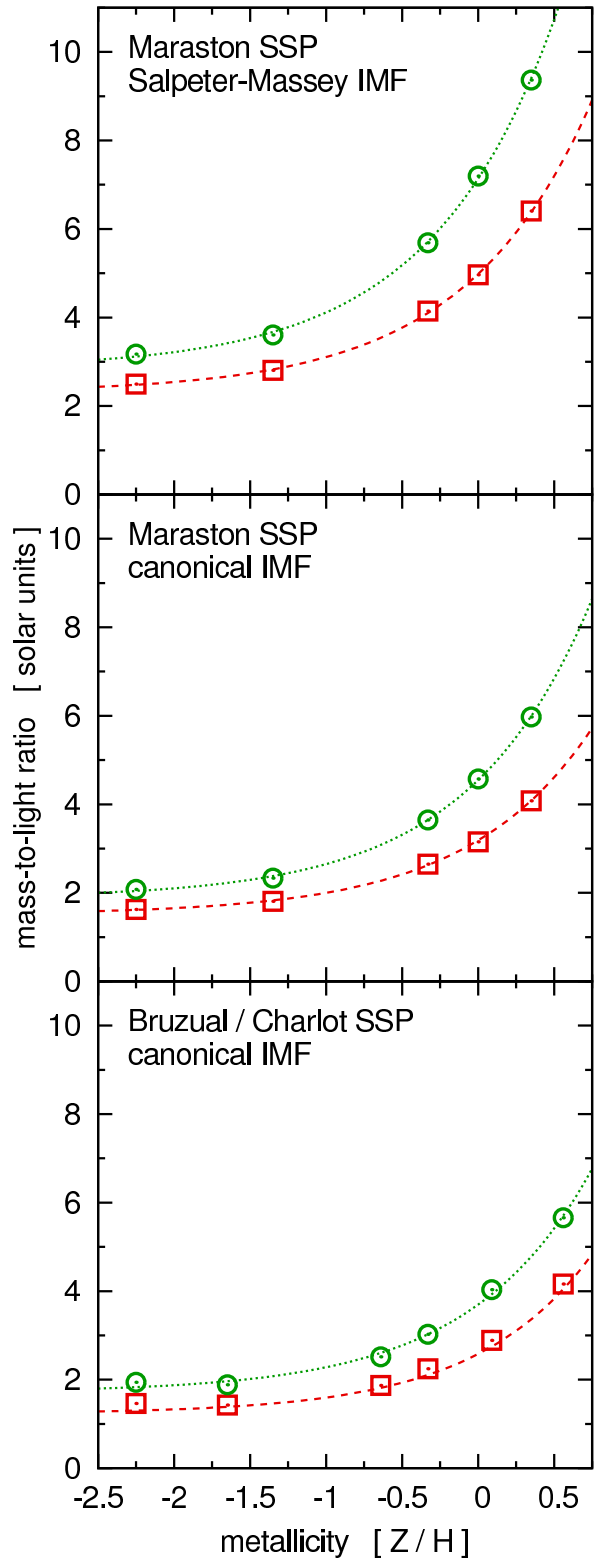}
\caption{The dependency of $\Upsilon_V$ on $Z$ for different SSPs. The origin and the IMF of the SSP model are detailed in in the upper part of each panel. The squares correspond to the models for $9 \, \rmn{Gyr}$ old populations and circles correspond to the models for $13 \, \rmn{Gyr}$ old SSPs. The lines indicate the interpolation (eq.~\ref{eqMetfit}) between the data from the SSP models.}
\label{fig9}
\end{figure}

In order to make statements on the $\Upsilon_V$ of objects with any $Z$, an interpolation formula that covers the whole $Z$-interval is needed. While it should be fairly simple, it should also closely fit the $M/L_{V}$ ratios that \citet{Bru2003} and \citet{Mar2005} find for specific metallicities. A function of the form
\begin{equation}
F_{i} ([Z/\rmn{H}]) = (a^{[Z/\rmn{H}]+b} + c)\frac{\rmn{M}_{\odot}}{\rmn{L}_{\odot}},
\label{eqMetfit}
\end{equation}
where the index $i$ distinguishes the different SSP models, fulfils these requirements well enough as Fig.~\ref{fig9} visualises. It can therefore safely be assumed that deviant estimates for $\Upsilon_{V}$ are not due to an inadequate interpolation formula, but due to incorrect assumptions on the stellar population in the MCOs or to a failure of the SSP models. The parameters $a$, $b$ and $c$ found in least-squares fits are given in Tab.~\ref{tabFit}. Comparing these parameters for different SSP models with the canonical IMF reveals that they do not only depend on the assumed age of the SSP, but also on whether the SSP models come from \citet{Bru2003} or \citet{Mar2005}. This results in noticeably lower expectations for the $M/L_{V}$ ratio from the SSP models from \citet{Bru2003}, if compared to an in terms of age and IMF identical model from \citet{Mar2005}. This proves the relevance of different stellar evolutionary models for the predictions from the SSP models.

It should be mentioned that the value of $\Upsilon_V$ for the highest metallicity was left out for the fit of eq.~(\ref{eqMetfit}) to the data from \citet{Mar2005}, because the omitted value was obtained by using a different stellar evolution model than for the other data from \citet{Mar2005}. Moreover, excluding it results into a much closer fit of $F_{i} ([Z/\rmn{H}])$ to the remaining data, which already cover the metallicity range of the MCOs and the MWGCs.

\begin{table}
\caption{Fit parameters for the interpolation formula for $\Upsilon_V$ to the data from the SSP models. The SSP models are from: 1: \citet{Mar2005}, 2: \citet{Bru2003}.}
\label{tabFit}
\centering
\vspace{2mm}
\begin{tabular}{lllll}
\hline
&&&&\\[-10pt]
Model & $a$ & $b$ & $c$ & Ref. \\ [2pt]
\hline
&&&&\\[-10pt]
Salpeter IMF, 9 Gyr & $3.33$ & $0.82$ & $2.30$ & 1\\
Salpeter IMF, 13 Gyr & $3.37$ & $1.20$ & $2.84$ & 1\\
canonical IMF, 9 Gyr & $3.42$ & $0.42$ & $1.51$ & 1\\
canonical IMF, 13 Gyr & $3.46$ & $0.79$ & $1.88$ & 1\\
canonical IMF, 9 Gyr & $3.70$ & $0.23$ & $1.23$ & 2\\
canonical IMF, 13 Gyr & $3.48$ & $0.55$ & $1.71$ & 2\\
\hline
\end{tabular}
\end{table}

Note that stellar evolution only raises the $M/L_{V}$ ratio of a stellar population. The $M/L_{V}$ ratio of a $13 \, \rmn{Gyr}$ old SSP therefore provides an upper limit for the $M/L_{V}$ ratio of a stellar population with a certain metallicity and IMF, since stellar populations cannot be much older according to the current estimates on the age of the universe ($13.73^{+0.16}_{-0.15} \, \rmn{Gyr}$; \citealt{Spe2007}).

If the stellar population of a star cluster with metallicity $Z_1$ is similar to one of the modelled SSPs, one would expect $\Upsilon_V$ to be close to the prediction from eq.~(\ref{eqMetfit}) for the $M/L_{V}$ ratio at $Z_1$:
\[
\Upsilon_{V} |_{Z_1} \approx F_i |_{Z_1}.
\]
If the stellar PDMF and the age of the star cluster is known (or assumed) to be similar to one of the SSP models that were introduced above and $Z_1$ has been measured, $\Upsilon_V$ of a cluster which has the metallicity $Z_2$, but is identical to the first one in all other respects can be estimated:
\begin{equation}
\Upsilon_{V} |_{Z_2} \approx \frac{F_i |_{Z_2}}{F_i |_{Z_1}} \times \Upsilon_{V} |_{Z_1}.
\label{eqMetNorm1}
\end{equation}
The division by $F_i |_{Z_1}$ is imposed by the condition that the estimate for $\Upsilon_V$ must not be changed for $Z_1 = Z_2$.  $\Upsilon_{V}|_{Z_1}/F_i|_{Z_1}$ is the factor by which the theoretical prediction for the $M/L_{V}$ ratio of a stellar system differs from the value that is observed. The multiplication of these numbers with $F_i|_{Z_2}$ is not necessary in principle, but it scales them by a constant $M/L_{V}$ ratio such that the predicted $M/L_{V}$ ratio from an SSP model with metallicity $Z_2$ is expected to coincide with an observed value, if the model is appropriate.

In order to eliminate the differences in $\Upsilon_V$ that are caused by differences in metallicity among the MCOs in the sample, we estimate $\Upsilon_V$ for them as it would be if they all had the same metallicity. This can be achieved by setting $Z_2$ identical for all objects while using the measured $Z$ for $Z_1$ in eq.~(\ref{eqMetNorm1}):
\begin{equation}
\Upsilon_{V \rmn{,n}}=\frac{F_i |_{\rmn{Z}_{\odot}}}{F_i |_{Z}} \times \Upsilon_{V},
\label{eqMetNorm2}
\end{equation}
where our (arbitrary) choice for $Z_2$ is the solar metallicity, $\rmn{Z}_{\odot}$.
We refer to the $M/L_{V}$ ratios calculated this way as the \textquotedblleft normalised $M/L_{V}$ ratios\textquotedblright, $\Upsilon_{V \rmn{,n}}$. Note that a comparison of a whole sample of values of observed $M/L_{V}$ ratios to a single prediction for the $M/L_{V}$ ratio of a SSP (as done in Fig. \ref{fig10}) becomes possible that way.

The values for $\Upsilon_{V \rmn{,n}}$ turn out to be quite insensitive to the actual choice out of the six sets of parameters $a$, $b$ and $c$ that encode different SSP models. This is due to the fact that the functions describing the dependency of $\Upsilon_V$ on $Z$ are almost identical up to a scale factor for all the model populations that are considered here, i.e. the ratio $F_i|_{Z_2}/F_i|_{Z_1}$ is almost independent of the SSP model chosen. This means that the $\Upsilon_{V \rmn{,n}}$ that are calculated here are very likely to be good representations of the $M/L$ ratios the MCOs and MWGCs would have if all their stars had solar composition, even if their PDMFs are different from all mass functions discussed here.

However, the choice of the SSP model certainly \emph{has} an impact on the prediction for the $M/L_{V}$ ratio of a population that completely fulfils the assumptions made for the model: For different models, the predictions on such a population would be different by about a factor of $F_i([Z/\rmn{H}])/F_j([Z/\rmn{H}])$.

\subsection{The normalised $M/L_{V}$ ratios of the MCOs and the MWGCs}

\begin{figure*}
\centering
\includegraphics[scale=0.85]{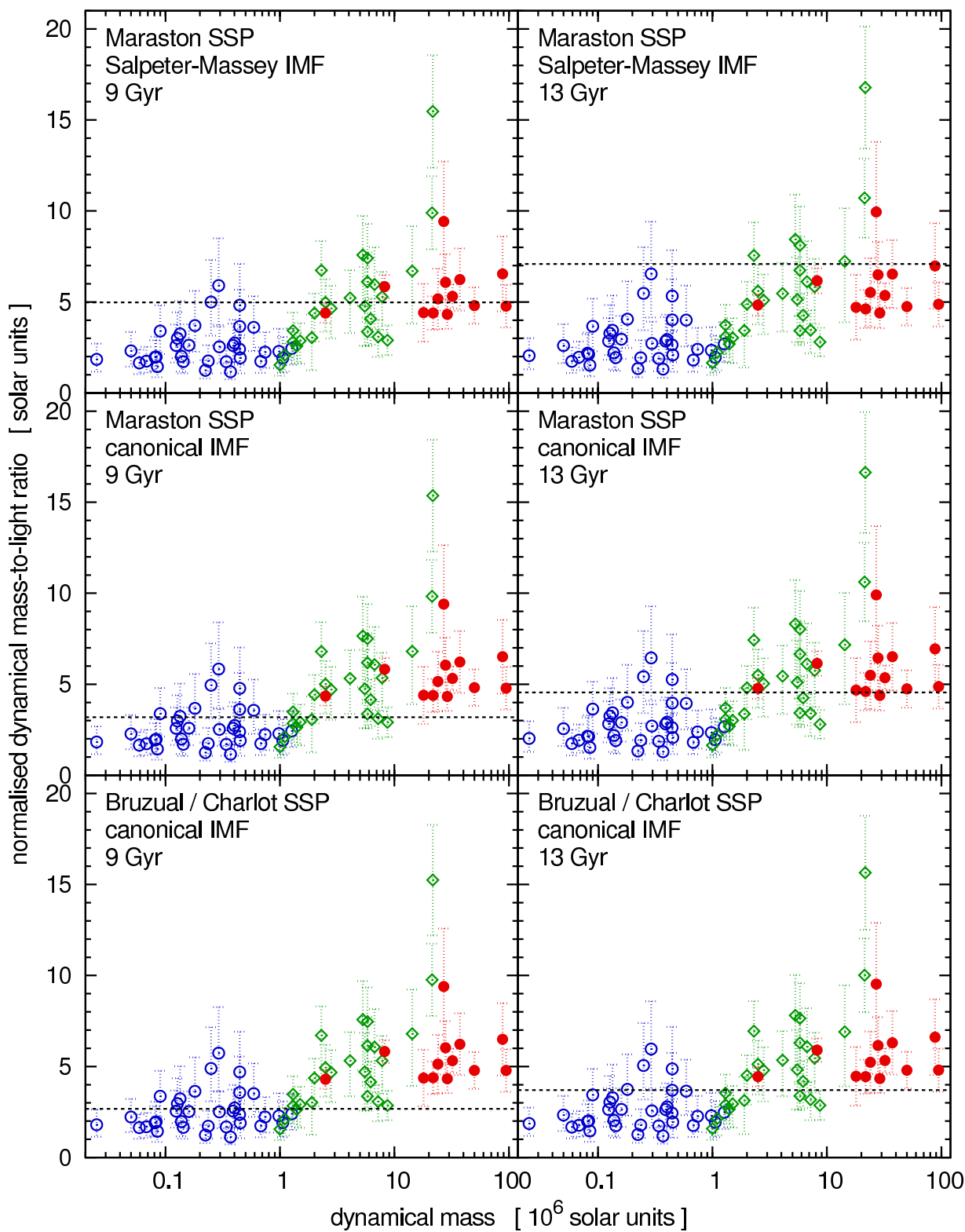}
\caption{Normalised mass-to-light ratio, $\Upsilon_{V\rmn{,n}}$, against mass for the MWGCs and MCOs based on the assumption that their stellar population can be described with SSP models. The origin of the SSP model, the assumed IMF and the assumed age of the SSP are given in the captions in each panel. The filled circles represent $\Upsilon_{V \rmn{,n}}$ of MCOs with measured $Z$, the open diamonds represent objects for which $Z$ was estimated from colour indices and open circles represent MWGCs. The dashed line indicates $F_i|_{\rmn{Z}_{\odot}}$, i.e. the $M/L_{V}$ ratio that the interpolation formula for the dependency of $\Upsilon_V$ on $Z$ predicts for $\rmn{Z}_{\odot}$, our reference metallicity. All points below that line have a lower $\Upsilon_V$ than the model predicts at their metallicity, all points above it exceed the model prediction. Naturally $F_i|_{\rmn{Z}_{\odot}}$ is very similar to the $M/L_{V}$ ratios at $\rmn{Z}_{\odot}$ given in the actual models, where such a direct comparison is possible. (\citet{Mar2005} have data on $\Upsilon_V$ for $\rmn{Z}_{\odot}$, \citet{Bru2003} use different grid points).}
\label{fig10}
\end{figure*}

The results for $\Upsilon_{V \rmn{,n}}$ of the MCOs and the MWGCs assuming different SSPs are presented in Fig.~\ref{fig10}.

The general distribution of the plotted points in all six panels of Fig.~\ref{fig10} still closely resembles the distribution of those points in Fig.~\ref{fig5}, which represent the same objects but with their observed $M/L_{V}$ ratios. However, the increase of the $M/L_{V}$ ratios from the MWGCs to the MCOs is less pronounced once the effect of the metallicity on the luminosity has been accounted for, since the metallicities of the MCOs are usually somewhat higher than the ones of MWGCs (Fig.~\ref{fig7}).

There is a large spectrum of values for the $\Upsilon_{V\rmn{,n}}$ of the MCOs, ranging from $\approx 2 \, \rmn{M}_{\odot} \, \rmn{L}_{\odot,V}^{-1}$ to $\approx 15 \, \rmn{M}_{\odot} \, \rmn{L}_{\odot,V}^{-1}$. However, most of them lie between $\approx 3 \, \rmn{M}_{\odot} \, \rmn{L}_{\odot,V}^{-1}$ and $\approx 7 \, \rmn{M}_{\odot} \, \rmn{L}_{\odot,V}^{-1}$. This still covers a large range of values, but taking into account that the $\Upsilon_{V\rmn{,n}}$ of individiual MCOs typically also are uncertain within a range of $\approx 2 \, \rmn{M}_{\odot} \, \rmn{L}_{\odot,V}^{-1}$ to $\approx 4 \, \rmn{M}_{\odot} \, \rmn{L}_{\odot,V}^{-1}$, it is not necessary to discuss physical reasons that could provide this scatter. However, two extreme outliers deserve more attention.

The first one of them is the faint MWGC NGC 6535, which has $\Upsilon_{V \rmn{,n}} \approx 15 \, \rmn{M}_{\odot} \, \rmn{L}_{\odot,V}^{-1}$ with large errors. As it is not only faint, but also fairly close to the galactic centre (the position is $l=27^{\circ} \, 18 '$, $b=10^{\circ} \, 44 '$ in Galactic coordinates), an accurate determination of its radius and velocity dispersion may be difficult due to the contamination with foreground stars. Moreover, its velocity dispersion has been derived from unpublished measurements. We therefore exclude it from Fig.~\ref{fig10}.

The second outlier is the MCO S999 in the Virgo cluster \citep{Has2005}, which is the object with the largest $\Upsilon_{V \rmn{,n}}$ in all panels of Fig.~\ref{fig10}. If this rather high value is not due to a flawed measurement, a scenario proposed by \citet{Fel2006} might offer an explanation. They proposed an enhancement of the $M/L_{V}$ ratio of MCOs by tidal interaction with the host galaxy. If this is indeed the case for S999, a faint envelope of stars may be detectable around it. It is noteworthy that this model can only provide an explanation for the $\Upsilon_{V\rmn{,n}}$ for a few MCOs out of a larger sample, as it requires quite specific orbital parameters.

A comparison of the predictions of the SSP models with solar metallicity with the values for calculated $\Upsilon_{V \rmn{,n}}$ shows that the bulk of MWGCs and MCOs with masses $\apprle 2 \times 10^6 \rmn{M}_{\odot}$ has lower $\Upsilon_{V \rmn{,n}}$ than it would be expected based on the assumed SSP models. Fig.~\ref{fig3} immediately reveals that these star clusters have relaxation times well below a Hubble time, which means that they are dynamically evolved due to their age. This result is therefore in (at least qualitative) agreement with the prediction by \citet{Bau2003b} and \citet{Bor2007}, who expect, based on their numerical simulations, the $M/L_{V}$ ratio of a star cluster in a tidal field to be lowered by dynamical evolution for most of its lifetime (See Section \ref{MassLight}).

The MCOs however have a strong tendency to \emph{higher} $M/L_{V}$ ratios compared to the theoretical prediction for a SSP with the canonical IMF, even for a $13 \, \rmn{Gyr}$ old population. There is only one SSP model, where in most of the cases the model expectation for $\Upsilon_{V\rmn{,n}}$ is higher than the actual $\Upsilon_{V\rmn{,n}}$ of the massive MCOs. This is the model with a $13 \, \rmn{Gyr}$ old stellar population which formed with a Salpeter-Massey IMF. For a $9 \, \rmn{Gyr}$ old population which formed with a Salpeter-Massey IMF, there seems to be agreement between the model prediction for $\Upsilon_V$ and the actual $\Upsilon_{V}$. However, assuming that the IMF is truly universal and recalling that the stellar PDMFs of MCOs should still reflect their stellar IMFs as their dynamical evolution is slow, it can be concluded that the stellar population of the MCOs should be well described by a SSP formed with the canonical IMF. The Salpeter-Massey IMF deviates in the low-mass part strongly from the canonical IMF and can thus be ruled out if the above assumptions hold.

It should be noted that the finding of observed $M/L_{V}$ ratios being higher than the theoretical prediction from a SSP model does not mean that the mass function of the chosen SSP model is inappropriate. Likewise, an agreement between the observed $M/L_{V}$ ratios and the prediction from the SSP model does not mean that the assumed IMF is correct. Consider for instance the presence of non-stellar black holes or non-baryonic DM in the MCOs, that lead to a rise of the $M/L_{V}$ ratio unaccounted for by any SSP model. However, in case the SSP model systematically overestimates the $M/L_{V}$ ratios of a sample of clusters, the model is certainly not a good description for the stellar population of the clusters.

Even if it is assumed that the MCOs only contain stars and stellar remnants, the significance of the tendency for higher $\Upsilon_{V \rmn{,n}}$ of the MCOs compared to SSPs whose IMFs agree with the canonical IMF should still be discussed. The case of a $13 \, \rmn{Gyr}$ old SSP with the canonical IMF from \citet{Mar2005} is of special interest and will therefore be treated in detail, because this is the model where the deviation of the $\Upsilon_{V \rmn{,n}}$ calculated for the MCOs from the theoretical expectation is the least pronounced. The values for $\Upsilon_{V \rmn{,n}}$ agree in fact with the prediction from the appropriate SSP model within the error for a large fraction of the MCOs, as can be seen in the middle right panel of Fig.~\ref{fig10}. On the other hand, if taken as a sample, the MCOs which are more massive than $2 \times 10^6 \, \rmn{M}_{\odot}$ still have a clear tendency for higher normalised $M/L_{V}$ ratios than one would expect from the SSP model.

A possibility to test whether a tendency is a significant deviation from an expectation is Pearson's test for the goodness of fit, as it is found in \citet{Bha1977} (see Appendix \ref{pearson}). We apply this test on the MCOs more massive than $2 \times 10^6 \, \rmn{M}_{\odot}$ under the assumption that their values for $\Upsilon_{V \rmn{,n}}$ would scatter just as much to higher values as to lower values compared to the prediction for $\Upsilon_{V \rmn{,n}}$ from an appropriate model.

The result of the test is then that the probability for the found (or an even more one-sided) distribution of the values for $\Upsilon_{V \rmn{,n}}$ of the MCOs more massive than $2\times 10^6 \, \rmn{M}_{\odot}$ around the expected value for a $13 \, \rmn{Gyr}$ old SSP with a canonical IMF from \citet{Mar2005} is $\ll 0.005$. The hypothesis that this SSP model can fully describe the properties of the MCOs can therefore be excluded according to this test.

The reliability of this result can be doubted, because it is not entirely clear whether the sample of the 31 objects, for which $M \ge 2 \times 10^6 \, \rmn{M}_{\odot}$ is fulfilled, is large enough to apply Pearson's test for the goodness of fit. Moreover, the objects with the more uncertain metallicity estimates from colour indices are included in this sample.

We therefore also apply the sign test, as described in \citet{Bha1977} (see Appendix \ref{sign}), on the 13 MCOs with metallicity estimates from line indices. The hypothesis to be tested is that there is no significant difference between their values for $\Upsilon_{V\rmn{,n}}$ and the theoretical expectation assuming a $13 \, \rmn{Gyr}$ old SSP with a canonical IMF from \citet{Mar2005}. The probability that the $\Upsilon_{V\rmn{,n}}$ are larger than the theoretical expectation in 12 or more cases is 0.002 according to this test, i.e. it is highly improbable that the hypothesis is correct.

\begin{figure}
\centering
\includegraphics[scale=0.80]{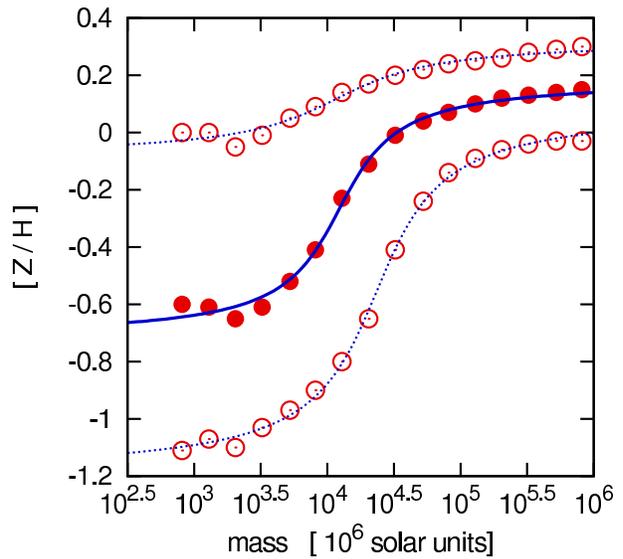}
\caption{The 16th percentiles (lower open circles), the median values (filled circles) and the 84th percentiles (upper open circles) of the distributions of the metallicities of galaxies in different total-stellar-mass bins \citep{Gal2005}. The lower dotted line, the solid line and the upper dotted line are our fits of eq.~(\ref{Galfit}) to the 16th percentiles, the median values and the 84th percentiles, respectively.}
\label{fig11}
\end{figure}

Both statistical tests thus suggest that stellar population models cannot explain the $M/L_{V}$ ratios as long as a canonical IMF is assumed, even for the maximum age the stellar population could have in order to be consistent with the age of the universe according to cosmological models. Note that \citet{Mie2006} suggest intermediate ages for the MCOs in the Fornax cluster. The actual discrepancy between the true values for $M/L_{V}$ ratios and the SSP models with the canonical IMF would then be larger than in the case discussed above. {\it This means that as long as the SSP models do not fail to describe real stellar populations, the MCOs either contain additional non-luminous matter, or their PDMFs must be different from what one would expect for a stellar system formed with the canonical IMF.}

\subsection{The normalised $M/L_{V}$ ratios of elliptical galaxies}

\begin{table}
\caption{Best-fitting parameters of eq.~(\ref{Galfit}) if fitted to the medians (50th percentiles) of the distributions of the metallicities of galaxies in different mass bins, as well as to the 16th and 84th percentiles of these distributions. The required data on the metallicity distributions is taken from \citet{Gal2005}, their table 2.}
\label{tabGalfit}
\centering
\vspace{2mm}
\begin{tabular}{lrrrr}
\hline
&&&&\\[-10pt]
Percentile & $a$ & $b$ & $c$ & $d$ \\ [2pt]
\hline
&&&&\\[-10pt]
Median (P50) & $0.29$ & $3.06$ & $-4.09$ & $-0.267$ \\
P16 & $0.41$ & $2.72$ & $-4.37$ & $-0.555$ \\
P84 & $0.13$ & $1.78$ & $-4.07$ & $0.118$ \\
\hline
\end{tabular}
\end{table}

\begin{figure*}
\centering
\includegraphics[scale=0.93]{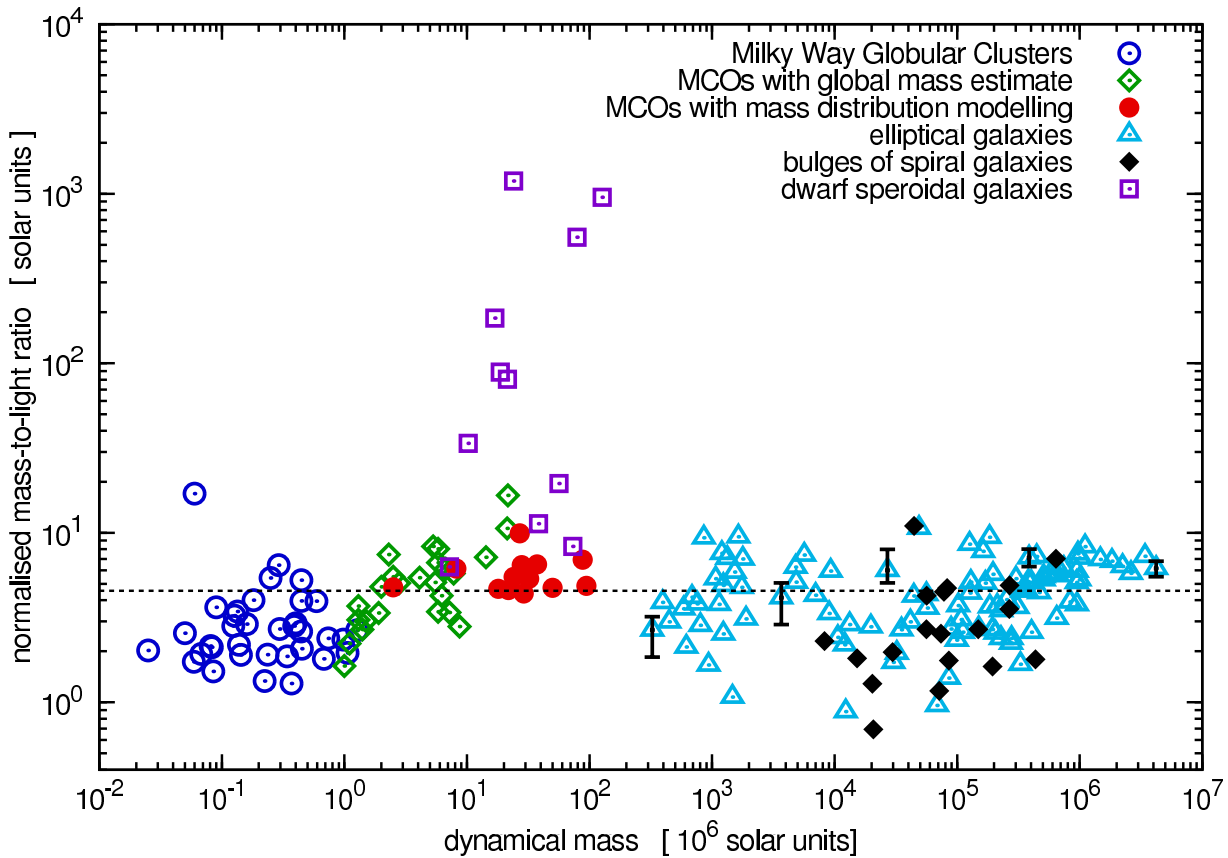}
\caption{The normalised $M/L_{V}$ ratios of all objects plotted in Fig.~\ref{fig5}. The symbols are as in Fig. 2. The stellar population model assumed for calculating the normalised $M/L_{V}$ ratios is the one from \citet{Mar2005} for a $13 \, \rmn{Gyr}$ old SSP (i.e. the same SSP model as for the middle right panel of Fig.~\ref{fig10}). Black bars indicate for five of the elliptical galaxies the range of normalised $M/L_{V}$ ratios they would assume if their metallicity would vary between the adopted values for the 16th percentile and the 84th percentile of the distribution of the metallicities of galaxies with that mass. The dashed line corresponds to the normalised $M/L_{V}$ ratio expected according to the SSP model assumed for the objects in this figure.}
\label{fig12}
\end{figure*}

We now compare the $M/L_{V}$ ratio of elliptical galaxies and galactic bulges with the prediction for the $M/L_{V}$ ratio of a $13 \, \rmn{Gyr}$ old SSP with the canonical IMF according to the models from \citet{Mar2005}.

The metallicity estimate that enters the calculation of the normalised $M/L_{V}$ ratio of the elliptical galaxies and galactic bulges is based on results on the metallicities of galaxies from the Sloan Digital Sky Survey obtained by \citet{Gal2005}. It is apparent from their data that the metallicities of galaxies in a given total-stellar-mass bin are distributed over a range of possible values (their figure~8 and table~2). In the present paper, the median of this distribution is taken as a representative value for the metallicities of the galaxies in that mass bin. The metallicities of the elliptical galaxies and galactic bulges in our sample as a function of their mass are calculated using the function
\begin{equation}
 [Z/\rmn{H}](M)=a \, \arctan \left( b \, \left[ \log \left( \frac{M}{10^6 \, \rmn{M}_{\odot}} \right) + c \right] \right) + d
\label{Galfit}
\end{equation}
with parameters $a$, $b$, $c$ and $d$ found in a least-squares fit to the median metallicities of galaxies in total-stellar-mass bins between $\approx 10^9 \, \rmn{M}_{\odot}$ and $\approx 10^{12} \, \rmn{M}_{\odot}$, as given by \citet{Gal2005}. The data from \citet{Gal2005} as well as the fit to them is shown in Fig.~\ref{fig11}. The best-fitting parameters $a$, $b$, $c$ and $d$ are noted in Tab.~\ref{tabGalfit}.

For the abundances of dSphs, it is assumed that their values for [Fe/H] can be identified with their values for [$Z$/H]. Iron abundances for most dSphs discussed here are given in \citet{Mat1998}, except for And II \citep{Con2005}, And XI \citep{Con2005} and UMa I \citep{Sim2007}.

The normalised $M/L$ ratios which are implied by the adopted metallicities for the elliptical galaxies, the galactic bulges and dSphs introduced in Section~\ref{Data} are plotted together with the normalised $M/L_{V}$ ratios of MCOs and MWGCs in Fig.~\ref{fig12}.

\citet{Gal2005} find especially for low-mass galaxies a large spread for the distribution of their metallicities. To quantify the uncertainties that arise for the adopted normalised $M/L_{V}$ ratios from the range of likely actual metallicities of galaxies, eq.~(\ref{Galfit}) is also fitted to the values from \citet{Gal2005} for the 16th and 84th percentiles of the distributions of metallicities of galaxies in different mass bins. The best fitting parameters $a$, $b$, $c$ and $d$ can be found in Tab.~\ref{tabGalfit}. Using these parameters, likely values for a high and a low metallicity in a given galaxy  can be estimated depending on its mass and the according normalised $M/L_{V}$ ratio can then be calculated. In Fig. \ref{fig12}, the possible range of normalised $M/L_{V}$ ratios suggested by the lower and the upper estimate of its metallicity is indicated for five sample objects with black bars.

It thereby becomes apparent in Fig. \ref{fig12} that the spread of the normalised $M/L_{V}$ ratios of elliptical galaxies and galactic bulges cannot be explained by different metallicities alone, but that at least one more parameter (e.g. the mean age of their stellar populations) must vary among them as well.

Consider the elliptical galaxies and galactic bulges with the highest normalised $M/L_{V}$ ratios. Given the adopted range for their likely metallicities, the range of $M/L_{V}$ ratios possible for them is inconsistent with the prediction for their normalised $M/L_{V}$ ratio from a model for a $13 \, \rmn{Gyr}$ old SSP from \citet{Mar2005}; especially for the objects with high dynamical masses. This suggests, as for the MCOs, an IMF different from the canonical IMF for their stellar populations or the presence of additional (gaseous or non-baryonic) matter in them.

\section{Discussion}
\label{Discussion}
\subsection{How reliable are the SSP models?}

\begin{figure*}
\centering
\includegraphics[scale=0.85]{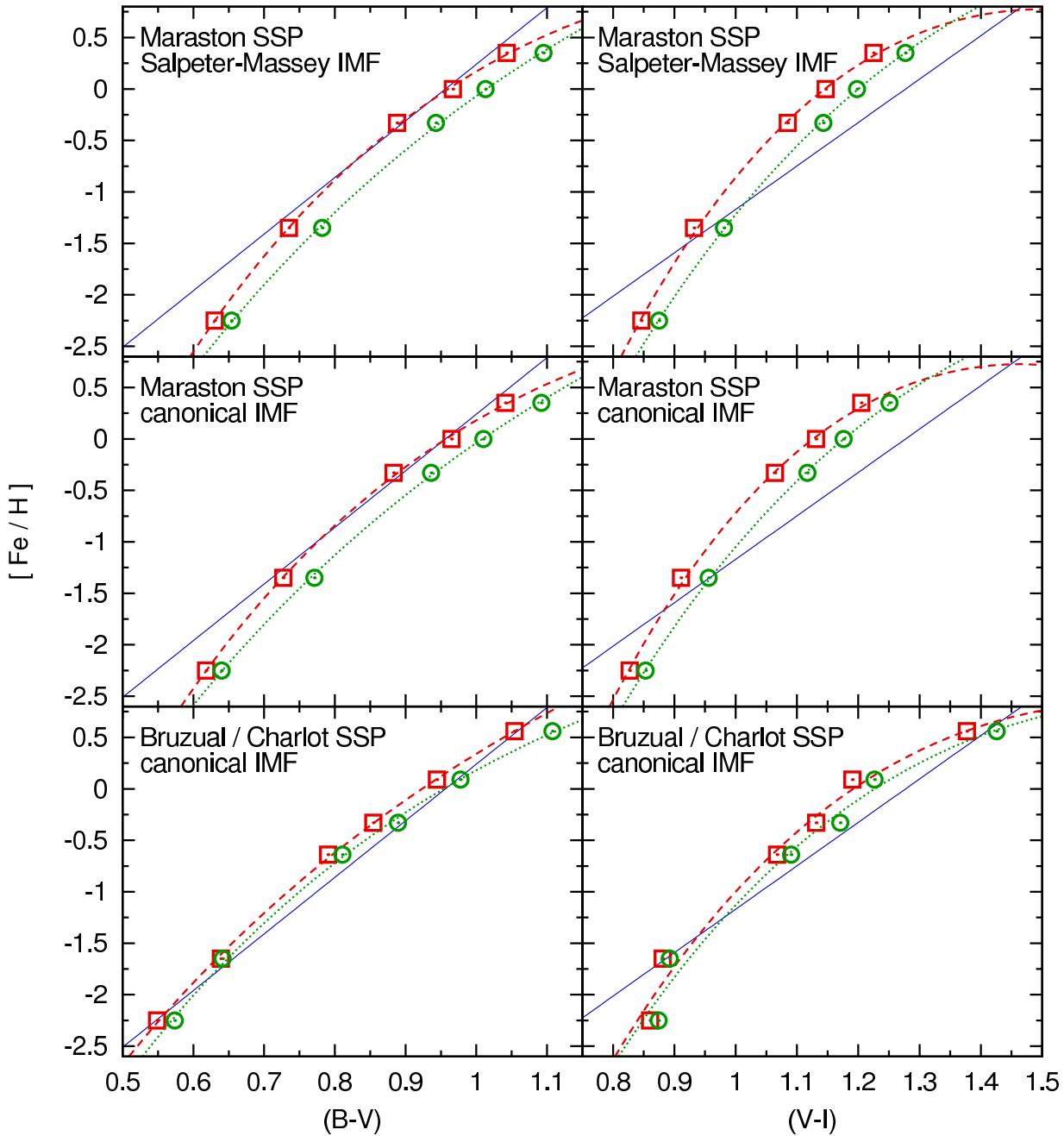}
\caption{The relation between colour indices and [Fe/H] according to SSP models. The right panels show [Fe/H] against the $(V-I)$ colour index while the left panels show [Fe/H] against the $(B-V)$ colour index. Squares show the data for $9 \, \rmn{Gyr}$ old populations. The dashed line is the fit to them. Circles show the data for $13 \, \rmn{Gyr}$ old populations. The dotted line is the fit to them. The thin solid lines represent the relations that have been established for MWGCs by \citet{Bar2000}. Out of the SSP models by \citet{Mar2005}, the case of a red horizontal branch is shown in this figure. This morphology is said to reflect the horizontal branches in most of the metal-rich GCs and therefore seems to be an appropriate choice for the MCOs, which show similar metallicities if measured.}
\label{fig13}
\end{figure*}

\begin{figure*}
\centering
\includegraphics[scale=0.85]{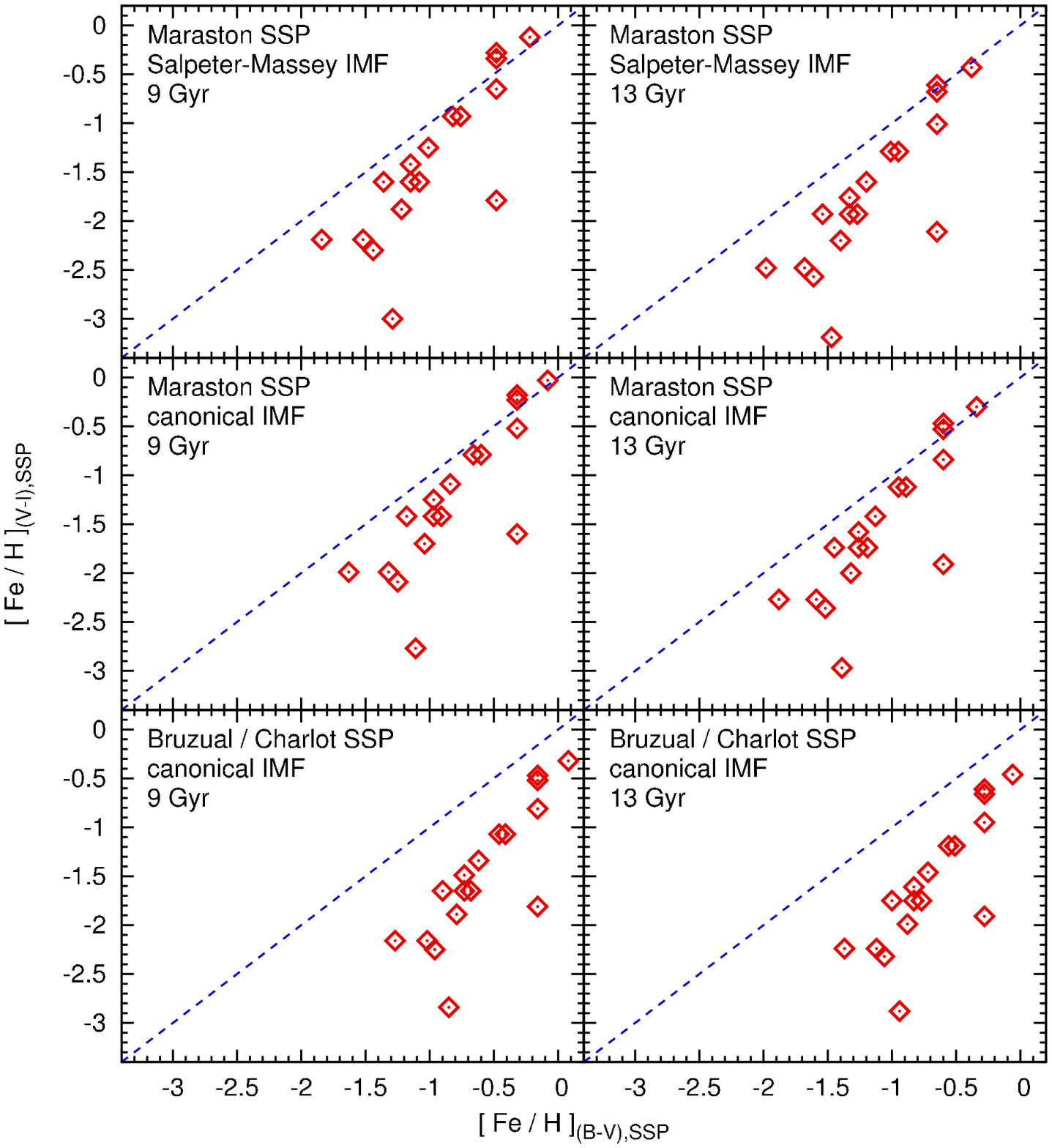}
\caption{A comparison between $[\rmn{Fe}/\rmn{H}]_{(B-V)\rmn{,SSP}}$ and $[\rmn{Fe}/\rmn{H}]_{(V-I)\rmn{,SSP}}$ for the objects in Centaurus~A (Tab.~\ref{tabCendata} and Tab.~\ref{tabCenmet}). The values are estimated by using the fits to the data from the SSP models plotted in Fig.~\ref{fig13}. The dashed line indicates equality of both estimates for the iron abundances. If there was no systematic difference between them for the objects in Centaurus~A, the distribution of the data would follow these lines. Errors to the plotted points are not shown. They are probably governed by the errors to the colour measurements (which are unknown to us) and by a mismatch between the SSP models and the real stellar populations of MCOs (which is to be shown by this figure), but not by the errors to the interpolations plotted in Fig. \ref{fig13}. For the SSP models from \citet{Mar2005}, a red horizontal branch is assumed. However, this does not have a strong impact on the distribution of the data in the according panels of this figure.}
\label{fig14}
\end{figure*}

The results that have been obtained in Section \ref{Met} are strongly based on the reliability of SSP models which are in turn based on the reliability of evolutionary stellar models. However, the reliability of these models cannot be taken for granted, as the differences between the models from \citet{Bru2003} and \citet{Mar2005} already indicate.

Another issue that may hint at difficulties with the SSP models is the relation between the iron abundance and the colour indices they suggest. This becomes apparent when using them to predict [Fe/H] of the MCOs in Centaurus~A from their colours. This can be done by setting up alternative equations to eqs.~(\ref{eqMet1}) and (\ref{eqMet2}) by fitting interpolation functions to the $(V-I)$-[Fe/H] value pairs and the $(B-V)$-[Fe/H] value pairs given by the SSP models (i.e. as in Section \ref{CompMet} for a relation between the metallicity and the $M/L_{V}$ ratio). Fig.~\ref{fig13} shows that a good fit between the data and the interpolation can be achieved with functions of the form
\begin{equation}
 [\rmn{Fe}/\rmn{H}]_{(V-I)\rmn{,SSP}}=a \, (V-I) +b(V-I)^{0.5}+c
\label{SSPMetColour}
\end{equation}
for the $(V-I)$ colour index and analogous for the $(B-V)$ colour index. The subscribt SSP in eq.~(\ref{SSPMetColour}) is supposed to indicate that these estimates for [Fe/H] from colour indices are based on SSP models, in contrast to the estimates for [Fe/H] from eqs.~(\ref{eqMet1}) and (\ref{eqMet2}), which are based on observations of the MWGCs.

In Fig.~\ref{fig14}, $[\rmn{Fe}/\rmn{H}]_{(V-I)\rmn{,SSP}}$ is plotted against $[\rmn{Fe}/\rmn{H}]_{(B-V)\rmn{,SSP}}$ for the objects in Centaurus~A. Each panel represents a choice of the SSP model which is assumed to represent the stellar population of the objects in Centaurus~A best. There are two features of the distribution of the data, which are remarkably little affected by that choice. The first one is the undeniable tendency for $[\rmn{Fe}/\rmn{H}]_{(V-I)\rmn{,SSP}}<[\rmn{Fe}/\rmn{H}]_{(B-V)\rmn{,SSP}}$. The second one is that the spread of the values for $[\rmn{Fe}/\rmn{H}]_{(V-I)\rmn{,SSP}}$ is larger than the spread of the values for $[\rmn{Fe}/\rmn{H}]_{(B-V)\rmn{,SSP}}$. However, if one of the SSP models is an adequate description for the actual SSPs in Centaurus~A, no systematic difference between the two estimates for [Fe/H] from this SSP model would be expected.

One could therefore come to the conclusion that none of the SSP models considered in this paper reflects the actual stellar populations of the objects in Centaurus~A. Note however that neither assuming an age of $5 \, \rmn{Gyr}$ nor considering a different horizontal branch morphology for the models from \citet{Mar2005} can enhance the concordance between $[\rmn{Fe}/\rmn{H}]_{(B-V)\rmn{,SSP}}$ and $[\rmn{Fe}/\rmn{H}]_{(V-I)\rmn{,SSP}}$ for the objects in Centaurus~A. This could be evidence of the standard SSP models failing to give a detailed and accurate description of real stellar populations in principle. \citet{Xin2007} claim that this might indeed be the case as long as SSP models are only based on the evolution of single stars but neglect the existence of blue stragglers, which are thought to be  products of stellar interactions. Given the complex abundance patterns in resolved massive star clusters, it also seems well possible that the observed (integrated) $(B-V)$ and $(V-I)$ color indices of the objects in Centaurus~A can only be reproduced by stellar population models which account for an age and metallicity spread of the stars.

An alternative explanation for the inconsistency between $[\rmn{Fe}/\rmn{H}]_{(B-V)\rmn{,SSP}}$ and $[\rmn{Fe}/\rmn{H}]_{(V-I)\rmn{,SSP}}$ could be a so far unidentified observational bias in the colour observations of the objects in Centaurus~A. This notion is made attractive by the finding that applying the observed relations eqs.~(\ref{eqMet1}) and (\ref{eqMet2}) for the estimation of [Fe/H] leads to smaller [Fe/H] estimates from $(V-I)$ colour indices than from $(B-V)$ colour indices for the MCOs in Centaurus~A as well (Fig. \ref{fig6}). If the difference between the metallicity estimates from eqs.~(\ref{eqMet1}) and (\ref{eqMet2}) was, for instance, caused by a systematic error to the $(B-V)$ colour indices, their offset from the true $(B-V)$ colour indices would be $\approx 0.1 \, \rmn{dex}$.

{\it Considering both the inconsistency of the iron abundances derived from the different colour indices by using the SSP models and the noticably different predictions of different SSP models on the $M/L_{V}$ ratio of the same population, it still seems possible that the enhancement of the $M/L$ ratios of the MCOs compared to the theoretical predictions for SSPs with the canonical IMF is due to a failure of the SSP models.}

\subsection{How reliable is an estimate of [Fe/H] from colour based on observations?}
\label{Uncertainties}

The alternative to estimating [Fe/H] from colour indices based on a SSP model is the approach chosen for this paper, namely using a relation between [Fe/H] and colour indices that has been established on a sample of observed star clusters, such as eqs.~(\ref{eqMet1}) and (\ref{eqMet2}). But just like the estimate of [Fe/H] by using SSP models, this approach is not unproblematic, as will be discussed here.

It is helpful to define two terms for the further discussion: We call the sample of objects for which the relation between [Fe/H] and colour was established the \textquotedblleft calibration sample\textquotedblright. The sample for which only colour indices are measured and where the relation between [Fe/H] and colour is used for a metallicity estimate is called the \textquotedblleft target sample\textquotedblright. In our specific case, the MCOs in Centaurus~A are the target sample and applying eqs.~(\ref{eqMet1}) and (\ref{eqMet2}) on them makes the MWGCs the calibration sample.

There are two problems, that are generally attached to an estimate of the iron abundances from the colours of objects in a target sample based on observations of an calibration sample. Firstly, it has to be assumed that the objects in both samples have at least typically the same PDMFs for shining stars and the same ages. If this is not the case, this method is likely to fail because colours depend on these parameters as well as on metallicity.

Secondly, relations such as eqs.~(\ref{eqMet1}) and (\ref{eqMet2}) are only \emph{fitting formulae} to a data sample with scatter. However, if these relations are applied to the objects in the calibration sample, the resulting estimates for [Fe/H] lie in the same parameter space as the values for [Fe/H] from line indices. The same is true if the calibration sample and the target sample are indeed comparable.

\begin{figure}
\centering
\includegraphics[scale=0.80]{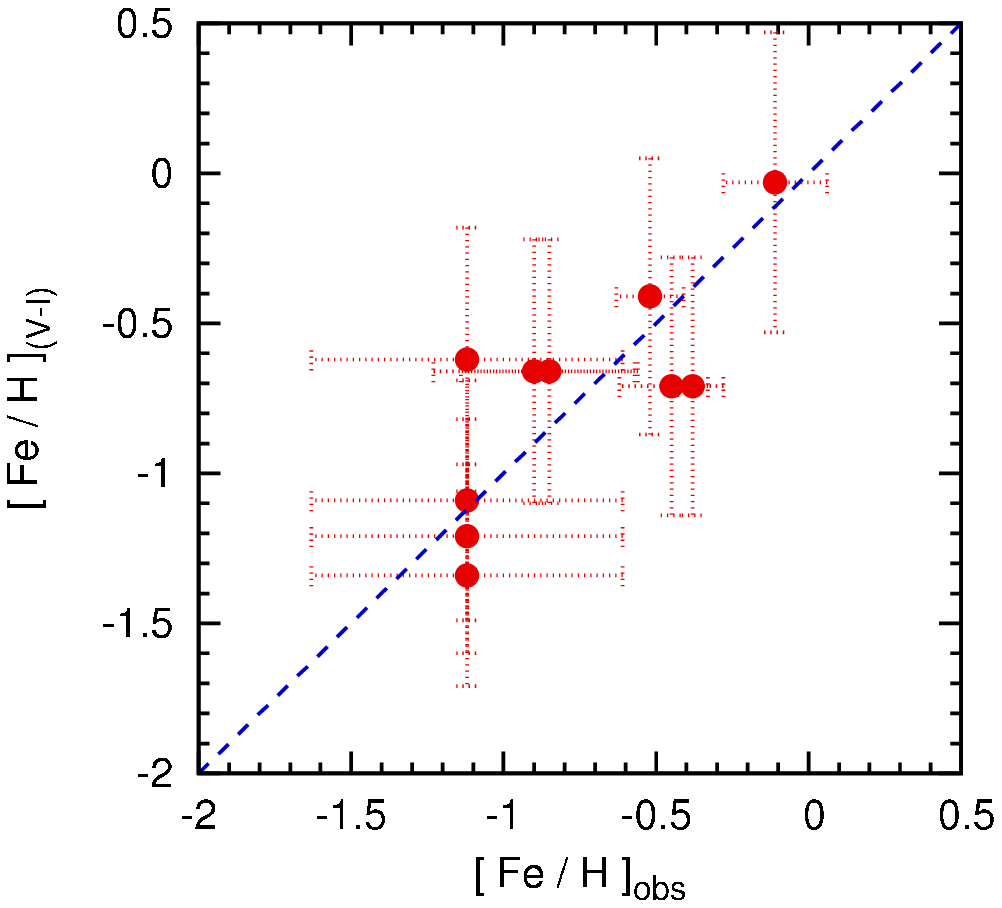}
\caption{$\rmn{[Fe/H]}_{(V-I)}$ calculated from eq.~\ref{eqMet1} plotted against the iron abundance estimate from line indices, $\rmn{[Fe/H]}_{\rmn{obs}}$, for the objects whose $(V-I)$ colours are given in Tab.~\ref{tabUCDmet}. The dashed line indicates equality between observed and calculated values. Apparently there is no clear tendency for the points to be located only on one side of the line. This indicates that there is no systematic difference between $\rmn{[Fe/H]}_{(V-I)}$ and $\rmn{[Fe/H]}_{\rmn{obs}}$.}
\label{fig15}
\end{figure}

As a test whether the MWGCs are a good choice for the calibration sample for the MCOs in Centaurus~A, the values for $\rmn{[Fe/H]}_{(V-I)}$ from eq.~(\ref{eqMet1}) are compared to the values for the estimates of the iron abundances from line indices (and thus directly linked to a observed iron content in the star clusters), $\rmn{[Fe/H]}_{\rmn{obs}}$. The published data (see Tab.~\ref{tabUCDmet}) allow such a comparison for the ten objects plotted in Fig.~\ref{fig15}.

There is no significant trend for $\rmn{[Fe/H]}_{(V-I)}$ to be larger or smaller than  $\rmn{[Fe/H]}_{\rmn{obs}}$, as the application of the sign test (\citealt{Bha1977}, Appendix \ref{sign}) shows. Under the hypothesis that there is no significant difference between the two values, the probability for having only four or less out of ten with $\rmn{[Fe/H]}_{\rmn{obs}} > \rmn{[Fe/H]}_{(V-I)}$ is 0.377. A result as the one plotted in Fig. \ref{fig15} is therefore quite probable. From this point of view it seems justifiable to apply eq.~(\ref{eqMet1}) on the MCOs, although it was originally fitted to the MWGCs.

Recall however that $\rmn{[Fe/H]}_{(B-V)}$ is systematically higher than $\rmn{[Fe/H]}_{(V-I)}$ for the objects in Centaurus~A (Fig.~\ref{fig6}). Since we adopt the mean of $\rmn{[Fe/H]}_{(V-I)}$ and $\rmn{[Fe/H]}_{(B-V)}$, [Fe/H] of the clusters in Centaurus~A will be overestimated if $\rmn{[Fe/H]}_{(V-I)}$ reflects their true abundances well. This is a conservative choice in our case, because a higher estimate for [Fe/H] leads to a lower estimate for $\Upsilon_{V \rmn{,n}}$. We arrived at the result that a SSP model with the canonical IMF underpredicts the $M/L_{V}$ ratios of the MCOs nevertheless, this therefore being a robust conclusion.

\subsection{The impact of a wrong estimate of [Fe/H] on the comparison of the dynamical $M/L_{V}$ ratios with the SSP models}

As the metallicities of the MCOs may be subject to systematic errors, it makes sense to discuss the impact of a wrong metallicity estimate on our claim that the $M/L_{V}$ ratios of the MCOs are inconsistent with the predictions from SSP models with the canonical IMF. We discuss one case in detail in order to give an impression how this affects our results.

Suppose the objects in the calibration sample are well described by a SSP with the same mass function, but that the target sample is younger than the calibration sample. The colour of the objects in the target sample is then bluer than it would be if they were of the same age as the objects in the calibration sample.

When relations like eqs.~(\ref{eqMet1}) and (\ref{eqMet2}) are applied in order to estimate the iron abundance, it is implicitly assumed that the stellar populations of the objects in the target sample are the same as the ones in the calibration sample. The estimates for the iron abundances are therefore too low if the target sample is younger than the calibration sample, because of the age-metallicity degeneracy \citep{Wor1994}. As a consequence, the $\Upsilon_{V \rmn{,n}}$ calculated from eq.~(\ref{eqMetNorm2}) is too high, since the denominator on the right side of eq.~(\ref{eqMetNorm2}) only decreases with decreasing $Z$ due to the exponential nature of eq.~(\ref{eqMetfit}).

However, the prediction for $\Upsilon_{V\rmn{,n}}$ made by a SSP model increases with the assumed age of the SSP. The expectation for the $\Upsilon_{V\rmn{,n}}$ of the objects in the target sample is therefore also too high, if they are compared to an SSP model which is, concerning the assumed age of the objects, more appropriate for the objects in the calibration sample. Thus, the error that is made in the estimation of the values for [$Z$/H] of the objects in the target sample by assuming a common age for all objects tends to balance the error that is made when all objects are compared to the same SSP model.

An analogous argument can be found if the objects in the target sample are depleted in low-mass stars compared to the objects in the calibration sample. In this case, it is the scarcity of low-mass stars that makes the objects in the target sample bluer and thereby leads to a too low metallicity estimate for them. The resulting too high estimate for $\Upsilon_{V \rmn{,n}}$ for these objects is compensated if they are compared to a SSP model with a full population of low-mass stars (which are faint and therefore enhance the $M/L_{V}$ ratio of the stellar population).

The reverse argumentation can be applied to objects with a higher age or more low-mass stars than the objects in the calibration sample.

It thereby seems that, also for objects with metallicity estimates from colour indices, finding the values for $\Upsilon_{V \rmn{,n}}$ above the expectation from the SSP model really is an indicator for additional non-luminous matter in the object.

\subsection{Implications of a high $M/L_V$-ratio in the MCOs}

Two explanations for the systematic enhancement of the $M/L_{V}$ ratios of the MCOs more massive than $2 \times 10^6 \, \rmn{M}_{\odot}$ compared to the predictions from SSP models with the canonical IMF are possible.

The first possibility is that the massive MCOs are embedded in DM haloes, as proposed by \citet{Has2005}. However, for MCOs with small effective radii and high $M/L_{V}$ ratios, the mean density of the DM within five half-light radii would have to be between $1 \, \rmn{M}_{\odot} \, \rmn{pc}^{-3}$ and $10 \, \rmn{M}_{\odot} \, \rmn{pc}^{-3}$ in order to have the observed impact on their dynamics. Adopting the universal DM density profiles as they are predicted by standard $\Lambda \, \rmn{CDM}$ cosmology \citep{Nav1996}, only DM haloes with masses of $10^{12} \, \rmn{M}_{\odot}$ or more could accumulate enough DM in their central parts (Dabringhausen \& Kroupa 2008a, in preparation).

The alternative to suggesting non-baryonic DM in the MCOs is to give up the notion of a universal IMF for all stellar populations. Such an alternative IMF would either be over-abundant in low-mass stars with high $M/L_{V}$ ratios (bottom-heavy IMF), see \citet{Mie2007}, or over-abundant in massive stars (top-heavy IMF). The latter possibility would imply a high number of dark stellar remnants in an old stellar population. Especially a top-heavy IMF seems attractive, since it is also suggested by models for galaxy evolution (e.g. \citealt{Bau2005,Nag2005,Dok2007}) or GC evolution (e.g. \citealt{Dan2004,Pra2006}). These issues will be examined in more detail in a forthcoming paper (Dabringhausen \& Kroupa 2008b, in preparation).

\subsection{On the nature of MCOs}

Apart from the finding that the MCOs more massive than $2 \times 10^6 \, \rmn{M}_{\odot}$ are in disagreement with the expectations for their $M/L_{V}$ ratios according to SSP models with the canonical IMF, the increase of typical radii at about the same mass is probably the most intriguing observation. This raises the question whether the massive MCOs (mostly classified as UCDs in the literature) constitute a population different to other populations of stellar systems as far as their origin is concerned. This question is of special interest for the relation between massive MCOs and GCs, since the seemingly \emph{continuous} rise of the mean radius above $10^6 \, \rmn{M}_{\odot}$ makes the notion of a single population of objects attractive (single population in the sense of a common scenario that leads to their formation). In this case, the evolution of such an object must be different at very high masses in order to account for the increase of radius with mass. A possible reason for this could be a dependency of star formation on gas density (which increases with mass for objects with the same extension, consider e.g. the MWGCs in Fig.~\ref{fig4}). If this dependency would lead to a greater mass loss during the lifetime of the cluster, it could explain the greater extensions because mass loss enlarges a star cluster. A possible example would be a top-heavy IMF in very dense star forming regions, which would cause a stronger mass loss by type II supernovae.

On the other hand, there have been efforts to design models that can specifically reproduce the parameters of UCDs. One of these models is the scenario by \citet{Fel2002} that UCDs are the merger of massive cluster complexes as are seen to be forming in massively interacting galaxies. Another one is the scenario by \citet{Bek2003} and \citet{Goe2007} that UCDs are the cores of nucleated galaxies\footnote{Note that the scenario \citet{Bek2003} proposes is inconsistent with $\Lambda \, \rmn{CDM}$ theory, because it has to assume that the DM haloes of the progenitors of the UCDs are cored instead of cusped.}. Under the condition that GCs form in the collapse of a single molecular cloud, objects that were formed in one of the above scenarios would indeed be of a different origin. This would offer natural explanations for the masses and the radii of those objects to be larger than for typical GCs. In this case, UCDs and GCs are two different populations that mix in the mass interval from $10^6 \, \rmn{M}_{\odot}$ to $10^7 \, \rmn{M}_{\odot}$ because both kinds of objects formed in intense starbursts that converted a similar amount of gas into stars.

A question connected to the issues discussed here is whether it is expedient to discriminate the MCOs into UCDs and GCs. We think that this distinction can be justified. It clearly makes sense if UCDs really formed in a different way than GCs. But it also makes sense in the case that GCs and UCDs were initially formed in the same way, thus are in principle to be considered as the same class of objects as far as their origin is concerned. In this case, \textquotedblleft UCD\textquotedblright would be a useful term to emphasise the peculiarities, for example the higher relaxation times, which very massive clusters usually show in comparison with their low-mass counterparts. Thus, UCDs could be \emph{defined} as those compact stellar systems, which have relaxation times longer than a Hubble time and thereby are (almost) collisionless systems on this time scale. This definition is the same as the one proposed by \citet{Kro1998} for a distinction between star clusters and galaxies, i.e. UCDs are galaxies in that sense.

Also G1 and $\omega$ Cen are classified as MCOs instead of GCs in this paper, because of the spread that their stars show in [Fe/H] ($\omega$ Cen certainly and G1 presumably). Peculiarities in element abundances can in principle be another way to discriminate UCDs from GCs by observational parameters. However, the MCOs in other galaxy clusters cannot be resolved into stars with the current instrumentation. A similar pattern of the chemical composition of their stars as in the MCOs in the Local Group can for this reason only be presumed so far, but not be proven in the near future.

\section{Conclusions}
\label{Conclusion}
In this paper, a sample of compact stellar systems covering the transition from globular clusters (GCs) to ultra compact dwarf galaxies (UCDs) and referred to as massive compact objects (MCOs) in this work, is compared to other dynamically hot stellar systems. Moreover, the $M/L_{V}$ ratios of the MCOs and the Milky Way GCs are compared to predictions from models for stellar populations. Our main conclusions are as follows.

Departing from radii typical for GCs, which are constant with mass, greater extensions are correlated with higher masses for dense stellar systems more massive than $10^6 \, \rmn{M}_{\odot}$. A strong increase of the median two-body relaxation time with mass is the natural consequence. We also find that stellar densities peak at a mass near $10^6 \, \rmn{M}_{\odot}$.

Dwarf spheroidal galaxies (dSphs) take on a special position among dynamically hot stellar systems. This is especially apparent from their dynamical $M/L_{V}$ ratios, which are in some cases higher by one to two orders of magnitude than for any other dynamically hot stellar system. Also note the large spread of the $M/L_{V}$ ratios of the dSphs, which would imply very different DM densities in the visible parts of different dSphs, if the dSphs were in dynamical equilibrium. It therefore seems improbable that the masses of dSphs can be determined by simple application of Jeans' equations.

The fact that compact stellar systems with $t_{rel} < \tau_{H}$ mostly have a much lower $M/L$ ratio than systems with $t_{rel} > \tau_{H}$ is qualitatively consistent with the findings by \citet{Bau2003b} and \citet{Bor2007}. However the change to the $M/L$ ratio by relaxation-driven mass loss does not exceed $0.3 \, \rmn{M}_{\odot} \, \rmn{L}_{\odot}^{-1}$ for most GCs, which is too little to explain the observed differences to the $M/L$ ratio of GCs and UCDs. Dynamical evolution is slow for UCDs, as their high relaxation times indicate, and consequently the decrease of the $M/L_{V}$ ratio by this process is slow as well. Moreover, the slow dynamical evolution leads to the stellar present-day mass function being almost identical with the stellar initial mass function for main sequence stars. We also found that the assumption of a population of old coeval stars in each massive MCO probably constitutes a good approximation to their real stellar populations.

Taken together, the lack of dynamical evolution and the narrow age spread of the stellar populations make a comparison between the MCOs and theoretical predictions from SSP models with widely used IMFs reasonable. The SSP models also allow to account for the differences due to the different metallicities of the MCOs. The limiting factor here is, if the reliability of the SSP models is taken for granted, the only rough knowledge of the element abundances in the MCOs. It turns out that the dynamical $M/L_{V}$ ratios of the MCOs more massive than $2 \times 10^6 \, \rmn{M}_{\odot}$ have a significant tendency to be even higher than the predictions of models for very old stellar populations, provided the IMF is chosen in agreement to the observations of stellar populations, where at present times low-mass main-sequence stars can be resolved (i.e. populations in the Milky Way and in objects in its immediate surroundings, such as the Magellanic Clouds).

It was shown however, that the SSP models that were used for the estimate of the expected $M/L_{V}$ ratio of the MCOs cannot produce consistent [Fe/H] estimates for the objects in Centaurus~A from the different colour indices measured for them. This poses the question whether the SSP models in their current state (e.g. without binary evolution) are truly reliable. On the other hand, if the predictions for the $M/L_{V}$ ratios from the SSP models are correct, the discrepancy between them and the dynamical $M/L_{V}$ ratios observed in the MCOs suggests that the more massive MCOs contain DM or that the stellar IMF in some stellar systems is different to the ones of resolved stellar populations. Both possibilities will be studied in follow-up papers.

Summarising, $\approx 10^6 \, \rmn{M}_{\odot}$ is a critical mass-scale at which the system length-scale begins to increase, the highest stellar density is reached, the relaxation time becomes comparable to a Hubble time and evidence for dark matter appears.

\section*{Acknowlegdements}
JD aknowledges support through DFG grant KR1635/13 and ESO funding for a one-week stay in Garching, where some of the ideas presented in this paper have been discussed. We thank M. Rejkuba for kindly making her data on the objects in Centaurus~A available to us before their publication. Fig.~\ref{fig8} has been prepared by J. Pflamm-Altenburg using a numerical routine described in the Appendix of \citet{Pfl2006}.

\bibliographystyle{mn2e}
\bibliography{paper11}

\appendix

\section[]{Statistical tests}

\subsection{Pearson's test for the goodness of fit}
\label{pearson}
Pearson's test for the goodness of fit \citep{Bha1977} can be used for deciding whether the frequency of a certain result for a measurement that has been performed on $n$ objects deviates significantly from an expected frequency. For the special case that only two results A and B can be the outcome of each measurement (A could be for example a result higher than a theoretical expectation and B the opposite case), result A will have occurred $j$ times and result B $n-j$ times. The probability of this outcome can now be calculated if a certain probability $p$ for the case A as the result of a measurement is assumed. A useful measure for this is given by the equation
\begin{equation}
\chi^2=\frac{(j-pn)^2}{pn}+\frac{((n-j)-(1-p)n)^2}{(1-p)n}.
\label{chi}
\end{equation}
There are tabulated values for $\chi^2$ (e.g. table~6 in the appendix of \citealt{Bha1977}) which make it possible to read off the probability for $\chi^2$ being higher than some value for a series of measurements, if the hypothesis for the probability $p$ is correct. (The degree of freedom is one in this case.)

\subsection{The sign test}
\label{sign}
The sign test \citep{Bha1977} is specifically designed for a small number of pairs of values, $(X_1,X_2)$, and is supposed to detect whether there is a significant trend for $X_2$ being larger or smaller than $X_1$ or not. $X_1$ and $X_2$ could be two measurements under different conditions (e.g. other instruments), or $X_1$ could be a value inferred from an observation, while $X_2$ is the theoretical prediction for this value. If the conditions under which $X_1$ and $X_2$ were obtained do not result into systematically larger or smaller values for $X_2$ compared to $X_1$, the probability for $X_1$ being larger than $X_2$ is 0.5. The probability that $X_2 > X_1$ for $j$ out of $n$ pairs of values is then given by the binomial distribution.

\label{lastpage}

\end{document}